\documentclass[twocolumn]{aastex631}

\usepackage{color}
\usepackage{natbib}
\usepackage{amsmath}
\usepackage{comment}

\newcommand{\nue}{\nu_{\rm e}}

\newcommand{\pa}{\partial}

\newcommand{\rd}{{\rm d}}

\shorttitle{}
\shortauthors{}
\graphicspath{{./}{figures/}}

\begin{document}

\title{Principal-Axis Analysis of the Eddington Tensor for the Early Post-Bounce Phase of Rotational Core-Collapse Supernovae}

\author[0000-0003-4959-069X]{Wakana Iwakami}
\affiliation{Advanced Research Institute for Science and Engineering, Waseda University, 3-4-1 Okubo, Shinjuku, Tokyo 169-8555, Japan}

\author[0000-0003-1409-0695]{Akira Harada}
\affil{Interdisciplinary Theoretical and Mathematical Sciences Program (iTHEMS), RIKEN, Wako, Saitama 351-0198, Japan}

\author[0000-0002-7205-6367]{Hiroki Nagakura}
\affiliation{Department of Astrophysical Sciences, Princeton University, Princeton, NJ 08544, USA}
\affiliation{Devision of Science, National Astronomical Observatory of Japan, 2-21-1, Osawa, Mitaka, Tokyo 181-8588, Japan}

\author[0000-0002-9234-813X]{Ryuichiro Akaho}
\affiliation{Advanced Research Institute for Science and Engineering, Waseda University, 3-4-1 Okubo, Shinjuku, Tokyo 169-8555, Japan}

\author{Hirotada Okawa}
\affiliation{Waseda Institute for Advanced Study, Waseda University, 1-6-1 Nishi-Waseda, Shinjuku-ku, Tokyo, 169-8050, Japan}

\author{Shun Furusawa}
\affiliation{Interdisciplinary Theoretical and Mathematical Sciences Program (iTHEMS), RIKEN, Wako, Saitama 351-0198, Japan}
\affiliation{College of Science and Engineering, Kanto Gakuin University, 1-50-1 Mutsuurahigashi, Kanazawa-ku, Yokohama, Kanagawa 236-8501, Japan}

\author{Hideo Matsufuru}
\affiliation{High Energy Accelerator Research Organization, 1-1 Oho, Tsukuba, Ibaraki 305-0801, Japan}

\author[0000-0002-9224-9449]{Kohsuke Sumiyoshi}
\affiliation{Numazu College of Technology, Ooka 3600, Numazu, Shizuoka 410-8501, Japan}

\author[0000-0002-2166-5605]{Shoichi Yamada}
\affiliation{Advanced Research Institute for Science and Engineering, Waseda University, 3-4-1 Okubo, Shinjuku, Tokyo 169-8555, Japan}



\begin{abstract}
Using full Boltzmann neutrino transport, we performed two-dimensional (2D) core-collapse supernova simulations in axisymmetry for two progenitor models with 11.2$M_{\odot}$ and 15.0$M_{\odot}$ both rotational and non-rotational.
We employed the results obtained in the early post-bounce phase ($t \lesssim 20$ ms) to assess performance under rapid rotation of some closure relations commonly employed in the truncated moment method.
We first made a comparison in 1D under spherical symmetry, though, of the Eddington factor $p$ defined in the fluid rest frame (FR).
We confirmed that the maximum entropy closure for the Fermionic distribution (MEFD) performs better than others near the proto-neutron star surface, where $p<1/3$ occurs, but does not work well even in 1D when the phase space occupancy satisfies $e < 0.5$ together with $p < 1/3$, the condition known to be not represented by MEFD.
For the 2D models with the rapid rotation, we employed the principal-axis analysis of the Eddington tensor.
We paid particular attention to the direction of the longest principal axis.
We observed in FR that it is aligned neither with the radial direction nor with the neutrino flux in 2D, particularly so in convective and/or rapidly rotating regions, the fact not accommodated in the moment method.
We repeated the same analysis in the laboratory frame (LB) and found again that the direction of the longest principal axis is not well reproduced by MEFD because the interpolation between the optically thick and thin limits is not very accurate in this frame.
\end{abstract}



\section{Introduction}

Core-collapse supernovae have been studied by many researchers over the years \citep[for reviews, see][]{Muller2016, Janka2016, Mezzacappa2020, Burrows2021}.
Although neutrino heating, which plays a crucial role in explosion, should be calculated as accurately as possible, various approximations have been used for multi-dimensional neutrino transfer to reduce the computational cost \citep[][]{OConnor2018, Vartanyan2019, Glas2019, Nagakura2020, Powell2020}.
One of the most commonly used is the two-moment method.
The 0th and 1st moments of the transport equation \citep[][]{Lindquist1966, Castor1972} are solved with a closure relation, in which the 2nd or 3rd moments are approximately given in terms of the lower-order moments \citep[][]{Shibata2011}.
This approach is called the M1 method \citep[][]{Murchikova2017} or the algebraic Eddington factor method \citep{Just2015}.
The Eddington factor $p$, which is used to interpolate between the optically thick and thin limits, is generally described as a function of the flux factor $f$.

Various functional forms of the Eddington factor have been proposed \citep[][]{Smit2000, Murchikova2017}.
The Wilson closure \citep{Wilson1975} and Kershaw closure \citep{Kershaw1976} were presented as flux limiters for the diffusion approximation.
The most popular closure is probably the Levermore closure \citep[][]{Levermore1984}.
The underlying idea of this closure is the following: if the stress-energy tensor of the isotropic radiation is transformed into other inertial frames by Lorentz transformations, then the relation between $p$ and $f$ is uniquely determined, irrespective of the chosen inertial frames: this relation is imposed as the closure.
The maximum entropy closures for Bose--Einstein radiation (MEBE) as well as for Fermi--Dirac radiation (MEFD) were studied by \citet[][]{Cernohorsky1994}.
Maximizing entropy for the Bose--Einstein or Fermi--Dirac statistics, the resultant angular distribution in momentum space gives the two-dimensional closure $p=p(f, e)$, where $e$ is called the number density or the phase space occupancy.
The well-known Minerbo closure \citep[ME,][]{Minerbo1978} can be regarded as the maximum entropy closure in the limit of $e \rightarrow 0$, in which both statistics give the same relation, while the Levermore--Pomraning closure \citep{Levermore1981} corresponds to the MEBE closure for $e \rightarrow \infty$.

The performance of these closures has been assessed by comparing them with data obtained by Monte Carlo simulations \citep[][]{Janka1992, Murchikova2017, Richers2017} or by simulations with the discrete-ordinate $(S_N)$ method \citep[][]{Smit2000, Richers2017, Nagakura2018, Harada2019, Harada2020, Iwakami2020} for matter distributions extracted from core-collapse simulations.
Most of the earlier papers employed spherically symmetric backgrounds, either taken from 1D simulations or angle-averaged for the results of 2D (axisymmetric) simulations.
\citet{Janka1992} demonstrated that the Fermi-Dirac form with two parameters that corresponds to the maximum entropy argument above gives a reasonable approximation.
\citet{Smit2000} also found for pre- and post bounce snapshots taken from a 1D simulation that the maximum entropy closure yields results  closest to the simulation data obtained with the $S_N$ method.
\citet{Murchikova2017} employed angle-averaged backgrounds at $t=160$, 260, and 360 ms after bounce in the 2D simulations by \citet{Ott2008} also with the $S_N$ method.
They found that no single closure works better than the others but that ME and MEFD closures yield better results more often than not.
These works commonly pointed out that there are regions of $p < 1/3$ in the supernova core but only the Wilson and MEFD closures can reproduce such situations; since the former does not satisfy the causality requirement \citep[see][]{Smit2000}, the latter was favored.

More recent investigations \citep{Richers2017,Nagakura2018,Harada2019,Harada2020,Iwakami2020} employed the results of multi-dimensional simulations as they are.
For example, \citet{Richers2017}  made a detailed comparison between their neutrino-transfer simulations with the Monte Carlo method or the discrete ordinate method for the same background model taken from a 2D simulation implemented with the latter scheme and elucidated pros and cons of each method.
They also assessed the performance of some closure relations.
\citet{Nagakura2018} and \citet{Harada2020}, on the other hand, studied the neutrino distributions in momentum space rather in detail, employing the results of their own 2D core-collapse simulations of non-rotating progenitors with the $S_N$ method.
In these three papers, the stress tensor itself or the Eddington tensor was compared component-wise with the Levermore closure.
\citet{Harada2019} extended the analysis of \citet{Nagakura2018} to slowly rotating models.
Very recently, \citet{Iwakami2020} performed 3D core-collapse simulations with the discrete ordinate method and compared the resultant neutrino distributions in momentum space with 1D and 2D counterparts.
They adopted the principal axis analysis of the Eddington tensor, the diagnostic first proposed in \citet{Harada2019} and employed also in this paper, to show rather poor performance of the Levermore closure for the hemispheric neutrino distribution.

In this paper, we extend the previous assessments of the closure relations in two directions.
In 1D we make a more systematic comparison of different closures, employing the results of our core-collapse supernova simulation with the $S_N$ method.
We confirm that those closures except for the MEFD closure are all in trouble at places with the Eddington factor $p < 1/3$.
This occurs both in the proto-neutron star \citep{Janka1992, Smit2000, Murchikova2017, Iwakami2020} and behind the prompt shock wave \citep{Janka1992, Iwakami2020} in the early post bounce phase.
The problem is extended even to the MEFD closure when the phase space occupancy becomes $e<0.5$ simultaneously.
In 2D, on the other hand, we study in detail the direction of the longest principal axis under rapid rotation.
Note that in the previous study with the principal axis analysis for slowly rotating models \citep{Harada2019}, only the length of the longest principal axis, the 2D counterpart of the Eddington factor in 1D, was considered; \citet{Iwakami2020} investigated the direction of the principal axis in their 3D simulation but rotation was ignored.
It is obvious, however, that in the genuinely non-spherical situations with rapid rotation, systematic misalignment of the longest principal axis from the radial direction becomes significant.
Note that, in the algebraic closures the principal axis is constrained to be aligned with the flux vector by construction in the fluid rest frame. The validity of this approximation under such rapid rotation has not been addressed so far and that is the main focus of this paper. We will also pay attention to the subtle difference from the choice of the frames to impose the closure relation.

This paper is organized as follows.
The basic equations and numerical setup are described in Section \ref{sec:numerical}, the closures considered in this paper are summarized in Section \ref{sec:closures}, and the principal-axis analysis of the Eddington tensor in Section \ref{sec:analysis}.
Then we present the 1D and 2D results in Sections \ref{sec:1d} and \ref{sec:2d}, respectively.
Finally, conclusions are given in Section \ref{sec:conclusion}.
Throughout the paper, the metric signature with $-\ +\ +\ +$ and the units $c=G=h=1$ are used unless otherwise stated, where $c$, $G$, and $h$ denote the speed of light, the gravitational constant, and Planck's constant, respectively.
The Greek $(\zeta, \eta, \xi)$ and Latin $(i, j, k)$ indices run over 0--3 and 1--3, respectively.

\section{Numerical Modeling \label{sec:numerical}}

\begin{figure*}[ht!]
\begin{center}
\includegraphics[width=\hsize]{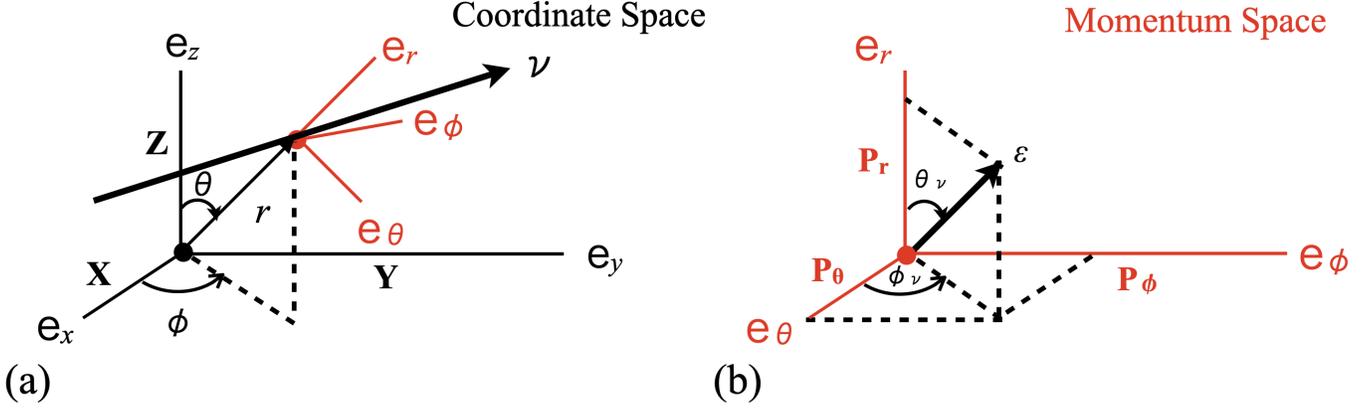}
\end{center}
\caption{The coordinate system in the phase space, where $r$, $\theta$, $\phi$, $\epsilon$, $\theta_\nu$, and $\phi_\nu$ are the radius, polar and azimuthal angles in space, the neutrino energy, and the polar and azimuthal angles in momentum space, respectively.
The three orthogonal axes in space are $\mathrm{X}$, $\mathrm{Y}$, and $\mathrm{Z}$ (a), and those in the momentum space are  $\mathrm{P_\theta}$, $\mathrm{P_\phi}$, and $\mathrm{P_r}$ (b).
The directions of the three orthogonal axes $\mathrm{P_\theta}$, $\mathrm{P_\phi}$, and $\mathrm{P_r}$ in momentum space respectively agree with those of the basis vectors $\mathbf{e}_\theta$,  $\mathbf{e}_\phi$, and $\mathbf{e}_r$ in space.
\label{fig:coordinate}}
\end{figure*}

The numerical method used in this study is essentially the same as that used by \citet{Nagakura2018} and \citet{Harada2019, Harada2020}.
The Boltzmann-radiation-hydrodynamics code based on the discrete ordinate $S_N$ method \citep{Sumiyoshi2012,Nagakura2014,Nagakura2017,Nagakura2019} is employed for core-collapse simulations.
The Boltzmann equation for neutrinos with a general metric in the conservative form \citep{Shibata2011} is:
\begin{eqnarray}
\frac{1}{\sqrt{-g}}\frac{\partial}{\partial x^\eta}\Bigg|_{q_j} \left[ \left( e_{(0)}^\eta + \sum_{j=1}^3 \ell_{(j)} e_{(j)}^\eta \right) \sqrt{-g}\mathcal{F}\right] && \nonumber\\
- \frac{1}{\epsilon^2}\frac{\partial}{\partial \epsilon}(\epsilon^3 \mathcal{F} \omega_{(0)}) + \frac{1}{\sin \theta_\nu} \frac{\partial}{\partial \theta_\nu}(\sin \theta_\nu \mathcal{F} \omega_{(\theta_\nu)} ) && \nonumber\\
+ \frac{1}{\sin^2 \theta_\nu}\frac{\partial}{\partial \phi_\nu}(\mathcal{F} \omega_{(\phi_\nu)}) = S_{\rm rad},
\end{eqnarray}
\begin{eqnarray}
\ell_{(j)} &=& (\cos \theta_\nu, \sin \theta_\nu \cos \phi_\nu, \sin \theta_\nu \sin \phi_\nu), \\
\omega_{(0)} &=& \epsilon^{-2}p^\eta p_\xi \nabla_\eta e_{(0)}^\xi, \\
\omega_{(\theta_\nu)} &=& \sum_{j=1}^3 \omega_j \frac{\partial \ell_{(j)}}{\partial \theta_\nu}, \\
\omega_{(\phi_\nu)} &=& \sum_{j=2}^3 \omega_j \frac{\partial \ell_{(j)}}{\partial \phi_\nu}, \\
\omega_j &=& \epsilon^{-2} p^\eta p_\xi \nabla_\eta e_{(j)}^\xi,
\end{eqnarray}
where $\mathcal{F}$, $g$, $x^\eta$, $p^\eta$, $S_{\rm rad}$, and $e^\xi_{(\eta)}$ are the distribution function, the determinant of the metric $g_{\eta\xi}$, the spatial coordinates, the four-momentum, the collision term, and the $\xi$ component of the tetrad basis $\mathbf{e}_{(\eta)}$, respectively.
The spacetime metric of the $3+1$ decomposition is written as: 
\begin{eqnarray}
\rd s^2 &=& g_{\eta\xi} \rd x^\eta \rd x^\xi \\
&=& -\alpha^2 \rd t^2 + \gamma_{jk}(\rd x^j + \beta^j \rd t)(\rd x^k + \beta^k \rd t),
\end{eqnarray}
where $t$, $\gamma_{\eta\xi} = g_{\eta\xi} + n_\eta n_\xi$, $n^\xi$ $= (1/\alpha, -\beta^j/\alpha)$, $\alpha$, and $\beta^j$ are time, the spatial metric, the normal vector to the spatial hypersurface with $t=\mathrm{const}$, the lapse function, and the shift vector, respectively.
The zeroth tetrad basis $e^\xi_{(0)}$ is chosen to agree with $n^\xi$.
In the spherical coordinates $(r,\theta,\phi)$, the other tetrad bases are described as:
\begin{eqnarray}
\mathbf{e}_{(1)} = \mathbf{e}_r &=& \gamma_{rr}^{-1/2} \pa_r, \label{eq:tet1} \\
\mathbf{e}_{(2)} = \mathbf{e}_\theta &=& -\frac{\gamma_{r\theta}}{\sqrt{\gamma_{rr}(\gamma_{rr}\gamma_{\theta\theta}-\gamma_{r\theta}^2)}}\pa_r \\
& & +\sqrt{\frac{\gamma_{rr}}{\gamma_{rr}\gamma_{\theta\theta}-\gamma_{r\theta}^2}}\pa_\theta, \label{eq:tet2} \\
\mathbf{e}_{(3)} = \mathbf{e}_\phi &=& \frac{\gamma^{r\phi}}{\sqrt{\gamma^{\phi\phi}}}\pa_r+\frac{\gamma^{\theta\phi}}{\sqrt{\gamma^{\phi\phi}}}\pa_\theta+\sqrt{\gamma^{\phi\phi}}\pa_\phi, \label{eq:tet3}
\end{eqnarray}
where $\pa_j$ are the coordinate bases of the vector.
The flat spacetime moving with the proto-neutron star in the velocity of $\beta^j$ is considered.
Then, $\mathbf{e}_r = \pa_r$, $\mathbf{e}_\theta = r^{-1} \pa_\theta$, and $\mathbf{e}_\phi =  (r\sin\theta)^{-1} \pa_\phi$.
The neutrino energy $\epsilon$ corresponds to the distance from the origin in momentum space, and the neutrino-propagating-direction angles $\theta_\nu$ and $\phi_\nu$ are defined as the angle from $\mathbf{e}_r$ and the angle from $\mathbf{e}_\theta$ on the $\mathbf{e}_\theta$--$\mathbf{e}_\phi$ plane, respectively, as shown in Fig.~\ref{fig:coordinate}.

Along with the Boltzmann equation for neutrino radiation transport, the Newtonian compressible hydrodynamics equations and the time-evolution equation of electron number density are solved as follows:
\begin{eqnarray}
\partial_t \mathbf{Q}+\partial_j \mathbf{U}^j=
\mathbf{W}_h + \mathbf{W}_i+\mathbf{W}_a, 
\end{eqnarray}
\begin{eqnarray}
\mathbf{Q}=\sqrt{g}\left(\begin{array}{c}
     \rho \\
     \rho v_r \\
     \rho v_\theta \\
     \rho v_\phi \\
     \mathcal{E} + \frac{1}{2}\rho v^2 \\
     \rho Y_e
\end{array}\right),
\end{eqnarray}
\begin{eqnarray}
\mathbf{U}^j=\sqrt{g}\left(\begin{array}{c}
     \rho v^j\\
     \rho v_r v^j + \mathcal{P} \delta^j_r\\
     \rho v_\theta v^j  + \mathcal{P} \delta^j_\theta\\
     \rho v_\phi v^j + \mathcal{P} \delta^j_\phi\\
     \left(\mathcal{E} + \mathcal{P} + \frac{1}{2}\rho v^2\right)v^j \\
     \rho Y_e v^j
\end{array}\right),
\end{eqnarray}
\begin{eqnarray}
\mathbf{W}_h=\sqrt{g}\left(\begin{array}{c}
     0 \\
     -\rho\partial_r \Psi 
     + \rho r (v^\theta)^2
     + \rho r \sin^2\theta(v^\phi)^2
     +\frac{2\mathcal{P}}{r}\\
     -\rho r^2\partial_\theta \Psi
     +\rho\sin\theta \cos\theta (v^\phi)^2
     +\frac{\mathcal{P}\cos\theta}{\sin\theta}\\
     -\rho \partial_\phi \Psi\\
     -\rho v^j \partial_j \Psi\\
     0
\end{array}\right),
\end{eqnarray}
\begin{eqnarray}
\mathbf{W}_i=\sqrt{g}\left(\begin{array}{c}
     0 \\
     -G_r \\
     -G_\theta \\
     -G_\phi \\
     G_t \\
     -\Gamma
\end{array}\right),
    \label{eq:hydro4}
\end{eqnarray}
\begin{eqnarray}
\mathbf{W}_a=\sqrt{g}\left(\begin{array}{c}
     0 \\
     -\rho \dot{\beta_r} \\
     -\rho \dot{\beta_\theta} \\
     -\rho \dot{\beta_\phi} \\
     -\rho v^j \dot{\beta_j} \\
     0
\end{array}\right),
    \label{eq:hydro5}
\end{eqnarray}
where $\rho$, $v_j$, $\mathcal{P}$, $\mathcal{E}$, $Y_e$, and $\delta^j_k$ are the matter density, velocity, pressure, internal energy density, electron fraction, and Kronecker's delta, respectively.
The Newtonian gravitational potential $\Psi$ is governed by the Poisson equation:
\begin{equation}
\Delta \Psi = 4\pi \rho.
\end{equation}

The exchange of momentum and energy between neutrino and matter fields in Eq.~(\ref{eq:hydro4}) is described by:
\begin{eqnarray}
G^\eta \equiv \sum_s G^\eta_s,\\
G^\eta_s \equiv \int p^\eta \epsilon S_{{\rm rad}(s)} \rd V_p,
\label{eq:interaction1}
\end{eqnarray}
\begin{eqnarray}
\Gamma \equiv \Gamma_{\nu_e}-\Gamma_{\bar{\nu}_e},\\
\Gamma_s \equiv m_u \int \epsilon S_{{\rm rad}(s)} \rd V_p,
\label{eq:interaction3}
\end{eqnarray}
where $m_u$, $S_{{\rm rad}(s)}$, $dV_p= \epsilon d\epsilon d\Omega_\nu$, and $d\Omega_\nu=\sin\theta_\nu d\theta_\nu d\phi_\nu$ denote the atomic mass unit, collision term of the neutrino species $s$, invariant volume in the momentum space, and differential solid angle in the momentum space, respectively.
The subscript $s$ represents three neutrino species: $\nu_e$,  $\bar{\nu}_e$, and $\nu_x$.
Detailed information of the collision terms is given by \citet{Sumiyoshi2012}.

The equation of state (EOS) based on the liquid drop model of the nuclei and the Skyrme-type interaction with incompressibility K = 220 MeV by \citet{Lattimer1991} is used here.
The 2D spatial domain in the range $0\le r\le5000$ km and $0\le \theta \le \pi$ is divided into $384 (r) \times 64 (\theta)$ or $384 (r) \times 128 (\theta)$ grid cells,
and the momentum space in the range $0 \le \epsilon \le$ 300 MeV  over the entire solid angle is discretized into $20 (\epsilon) \times 10 (\theta_\nu) \times 6 (\phi_\nu)$ grid points.
The 11.2$M_\odot$ and 15.0$M_\odot$ progenitor models evolved without rotation \citep{Woosley2002} are employed in this study.
From the onset of the core collapse, 1D simulations are performed for non-rotating models, and 2D low-resolution simulations  are done for rotating models where a rigid rotation of 6 rad/s at $r=1000$ km is manually added initially.
When an entropy gradient becomes negative for the first time soon after the core bounce, the zenith grid number is switched to 128 and 0.1\% random velocity perturbations are imposed for both models.

\section{CLOSURES \label{sec:closures}}

The following explains how to calculate the Eddington factor used for the analyses in this paper.
An unprojected second moment \citep[][]{Thorne1981,Shibata2011} is  defined in an arbitrary frame as
\begin{eqnarray}
M^{\eta\xi}(\epsilon_\mathrm{FR}) \equiv \int \mathcal{F}\  \delta\left(\frac{\epsilon_\mathrm{FR}^3}{3}-\frac{(-u_\zeta p^\zeta)^3}{3}\right)p^\eta p^\xi dV_p,
\label{eq:moment}
\end{eqnarray}
where $u^\eta$ and $\epsilon_\mathrm{FR}$ are the four-velocity of medium and the neutrino energy measured in the fluid rest frame (FR), respectively.
The second angular moment can be written as
\begin{eqnarray}
M^{\eta\xi}(\epsilon_\mathrm{FR}) &= J(\epsilon_\mathrm{FR})u^{\eta}u^{\xi}+H^\eta(\epsilon_\mathrm{FR}) u^\xi \nonumber \\
&+ H^\xi(\epsilon_\mathrm{FR}) u^\eta + L^{\eta\xi}(\epsilon_\mathrm{FR}),
\label{eq:moment2}
\end{eqnarray}
where the energy density $J$, energy flux $H^\eta$, and radiation pressure tensor $L^{\eta\xi}$ are the variables projected on to the FR.

The terms $J$ and $H^\eta$ in Eq.(\ref{eq:moment2}) can be used to calculate the dimensionless parameters, that is, the phase space occupancy:
\begin{eqnarray}
e = \frac{J(\epsilon_\mathrm{FR})}{4\pi\epsilon_\mathrm{FR}},
\label{eq:e}
\end{eqnarray}
and the flux factor:
\begin{eqnarray}
f = \sqrt{\frac{h_{\eta\xi}H^{\eta}(\epsilon_\mathrm{FR})H^{\xi}(\epsilon_\mathrm{FR})}{J^2(\epsilon_\mathrm{FR})}},
\label{eq:f}
\end{eqnarray}
where ${h}_{\eta\xi}=g_{\eta\xi}+u_\eta u_\xi$ is the projection operator onto the FR.
In 1D, not the Eddington tensor but the Eddington factor, i.e., the ratio of the $rr$-component of the radiation pressure tensor to the energy density defined as
\begin{eqnarray}
p  = \frac{L^{rr}(\epsilon_\mathrm{FR})}{J(\epsilon_\mathrm{FR})}.
\end{eqnarray}
is used.
It is straightforward to calcualte it in the Boltzmann-transport simulation whereas it is prescribed algebraically in the truncated moment method.
The closure relations that are compared in this paper are as follows:
\begin{itemize}
\item[-] MEFD closure \citep{Cernohorsky1994}:
\begin{eqnarray}
p_\mathrm{\ MEFD} = \frac{1}{3}+\frac{2}{3}(1-e)(1-2e)\chi \left(\frac{f}{1-e}\right), \nonumber \\
\chi(x)=x^2(3-x+3x^2)/5; \ \
\label{eq:p_MEFD}
\end{eqnarray}
\item[-] the maximal packing envelope (MP) \citep{Smit2000}:
\begin{eqnarray}
p_\mathrm{\ MP}\ = \frac{1}{3}(1-2f+4f^2);
\label{eq:p_MEFDMP}
\end{eqnarray}
\item[-] ME closure in the classic limit \citep{Minerbo1978}:,
\begin{eqnarray}
p_\mathrm{\ ME}\ (e \rightarrow 0)= \frac{1}{3}+\frac{2f^2}{15}(3-f+3f^2);
\label{eq:p_ME}
\end{eqnarray}
\item[-] MEBE closure in the limit of $e\rightarrow\infty$ \citep{Levermore1981}:
\begin{eqnarray}
p_\mathrm{\ MEBE}\ (e \rightarrow \infty)= f \coth R, \nonumber \\
f = \coth R - 1/R;
\label{eq:p_MEBE1}
\end{eqnarray}
\item[-] the Levermore closure \citep{Levermore1984}:
\begin{eqnarray}
p_\mathrm{\ Levermore} = \frac{3+4f^2}{5+2\sqrt{4-3f}}.
    \label{eq:p_Levermore}
\end{eqnarray}
\end{itemize}

\section{PRINCIPAL-AXIS ANALYSIS OF THE EDDINGTON TENSOR \label{sec:analysis}}

In 2D, we employ the entire Eddington tensor.
This section briefly introduces the principal-axis analysis of the Eddington tensor \citep[see][for a detailed explanation]{Iwakami2020}.
The aforementioned closures are constructed with the assumption of the axisymmetric neutrino distribution around the energy flux vector in the momentum space, which is normally realized for spherically symmetric matter distributions in space.
The core-collapse supernovae are not spherically symmetric, though, owing to hydrodynamic instabilities or stellar rotation (or magnetic fields).
In fact, the neutrino angular distribution in momentum space is complicated in the convective flow in general, with the longest principal axis of the Eddington tensor not always parallel to the flux \citep{Iwakami2020}.
Note that, as shown below, the closure relation in the truncated moment method normally employs the flux vector alone to construct the tensor structure in the anisotropic part of Eddington tensor, thus forcing its longest principal axis either parallel or perpendicular to the flux.
The main goal of this paper is to assess this restriction in the closure relations in the genuinely non-spherical settings under rapid rotation.
In the following we adopt the MEFD closure as the canonical one, since it performs best according to our comparison in 1D as shown below.

The Eddington tensor calculated in the FR with the Boltzmann transport simulations:
\begin{eqnarray}
k^{ij}_\mathrm{Boltz} (\epsilon_\mathrm{FR})=\frac{L^{ij} (\epsilon_\mathrm{FR})}{J(\epsilon_\mathrm{FR})}, \label{eq:kijBoltzmann_FR}
\end{eqnarray}
is compared with the counterpart for the MEFD closure,
\begin{eqnarray}
k^{ij}_\mathrm{MEFD} (\epsilon_\mathrm{FR})=\frac{L^{ij}_{MEFD} (\epsilon_\mathrm{FR})}{J(\epsilon_\mathrm{FR})}. \label{eq:kijM1_FR}
\end{eqnarray}
The radiation pressure tensor for the MEFD closure in the FR is expressed as
\begin{eqnarray}
L^{ij}_{\rm MEFD} (\epsilon_\mathrm{FR}) = 
\frac{3}{2}(1-p_\mathrm{MEFD}) L^{ij}_{\rm thick} (\epsilon_\mathrm{FR}) \nonumber \\
+ \frac{1}{2}(3p_\mathrm{MEFD} - 1) L^{ij}_{\rm thin} (\epsilon_\mathrm{FR}),
\label{eq:Pijm1_FR}
\end{eqnarray}
\begin{eqnarray}
L^{ij}_{\rm thick} (\epsilon_\mathrm{FR}) = \frac{1}{3}J (\epsilon_\mathrm{FR})h^{ij},
\label{eq:Pthick_FR}
\end{eqnarray}
\begin{eqnarray}
L^{ij}_{\rm thin}(\epsilon_\mathrm{FR}) =J (\epsilon_\mathrm{FR}) \frac{H^i(\epsilon_\mathrm{FR})H^j(\epsilon_\mathrm{FR})}{H(\epsilon_\mathrm{FR})^2},
\label{eq:Pthin_FR}
\end{eqnarray}
where $L^{ij}_\mathrm{thick}$ and $L^{ij}_\mathrm{thin}$ are the radiation pressure tensors at the optically thick and thin limits, respectively, in the FR.

By contrast, in the laboratory frame (LB), the second angular moment can be written as
\begin{eqnarray}
M^{\eta\xi}(\epsilon_\mathrm{FR}) =
E(\epsilon_\mathrm{FR})n^{\eta}n^{\xi}+F^\eta(\epsilon_\mathrm{FR}) n^\xi \nonumber
\\ + F^\xi(\epsilon_\mathrm{FR}) n^\eta + P^{\eta\xi}(\epsilon_\mathrm{FR}),\label{eq:moment3}
\end{eqnarray}
where $E$, $F^{\eta}$, and $P^{\eta\xi}$ are the energy density, energy flux, and pressure tensor projected on to the LB, respectively.
Note that these variables are functions of the energy defined in the FR \citep{Shibata2011}.
The Eddington tensors are defined as
\begin{eqnarray}
k^{ij}_\mathrm{Boltz} (\epsilon_\mathrm{FR})=\frac{P^{ij} (\epsilon_\mathrm{FR})}{E(\epsilon_\mathrm{FR})} \label{eq:kijBoltzmann_LB}
\end{eqnarray}
for the Boltzmann transport and as
\begin{eqnarray}
k^{ij}_\mathrm{MEFD} (\epsilon_\mathrm{FR})=\frac{P^{ij}_{MEFD} (\epsilon_\mathrm{FR})}{E (\epsilon_\mathrm{FR})} \label{eq:kijM1_LB}
\end{eqnarray}
for the MEFD closure.
For the latter the radiation pressure tensor is given by
\begin{eqnarray}
P^{ij}_{\rm MEFD} (\epsilon_\mathrm{FR}) = 
\frac{3}{2}(1-p_\mathrm{MEFD}) P^{ij}_{\rm thick} (\epsilon_\mathrm{FR}) \nonumber \\
+ \frac{1}{2}(3p_\mathrm{MEFD} - 1) P^{ij}_{\rm thin} (\epsilon_\mathrm{FR}),
\label{eq:Pijm1}
\end{eqnarray}
\begin{eqnarray}
P^{ij}_{\rm thick} (\epsilon_\mathrm{FR}) =  \frac{1}{3}J(\epsilon_\mathrm{FR})\gamma^{ij}+ \frac{4}{3}J(\epsilon_\mathrm{FR}) v^i v^j \nonumber \\ + (H^i(\epsilon_\mathrm{FR}) v^j + v^i H^j(\epsilon_\mathrm{FR})),
\label{eq:Pthick}
\end{eqnarray}
\begin{equation}
P^{ij}_{\rm thin}(\epsilon_\mathrm{FR}) = E(\epsilon_\mathrm{FR}) \frac{F^i(\epsilon_\mathrm{FR})F^j(\epsilon_\mathrm{FR})}{F(\epsilon_\mathrm{FR})^2}, \label{eq:Pthin}
\end{equation}
where $P_\mathrm{thick}^{ij}$ and $P_\mathrm{thin}^{ij}$ are the radiation pressure tensors at the optically thick and thin limits, respectively, in the LB.
Equations~(\ref{eq:Pijm1}--\ref{eq:Pthin}) are directly evaluated from $E$, $F^i$, $J$, and $H^i$ obtained by taking the moments in the different frames in this paper. 
In the simulation with the truncated moment method, one normally knows the energy density and flux in either LB or FR.
Then, some iteration procedure is required to obtain the quantities in the other frame in a self-consistent way. 
However, this iteration procedure is not taken here as we have the second moment tensor $M^{\eta\xi}$, from which these quantities are obtained in both frames directly. 
\citet{Harada2020} showed that the two methods give negligible differences.

\begin{figure}[b]
\begin{center}
\includegraphics[width=8cm]{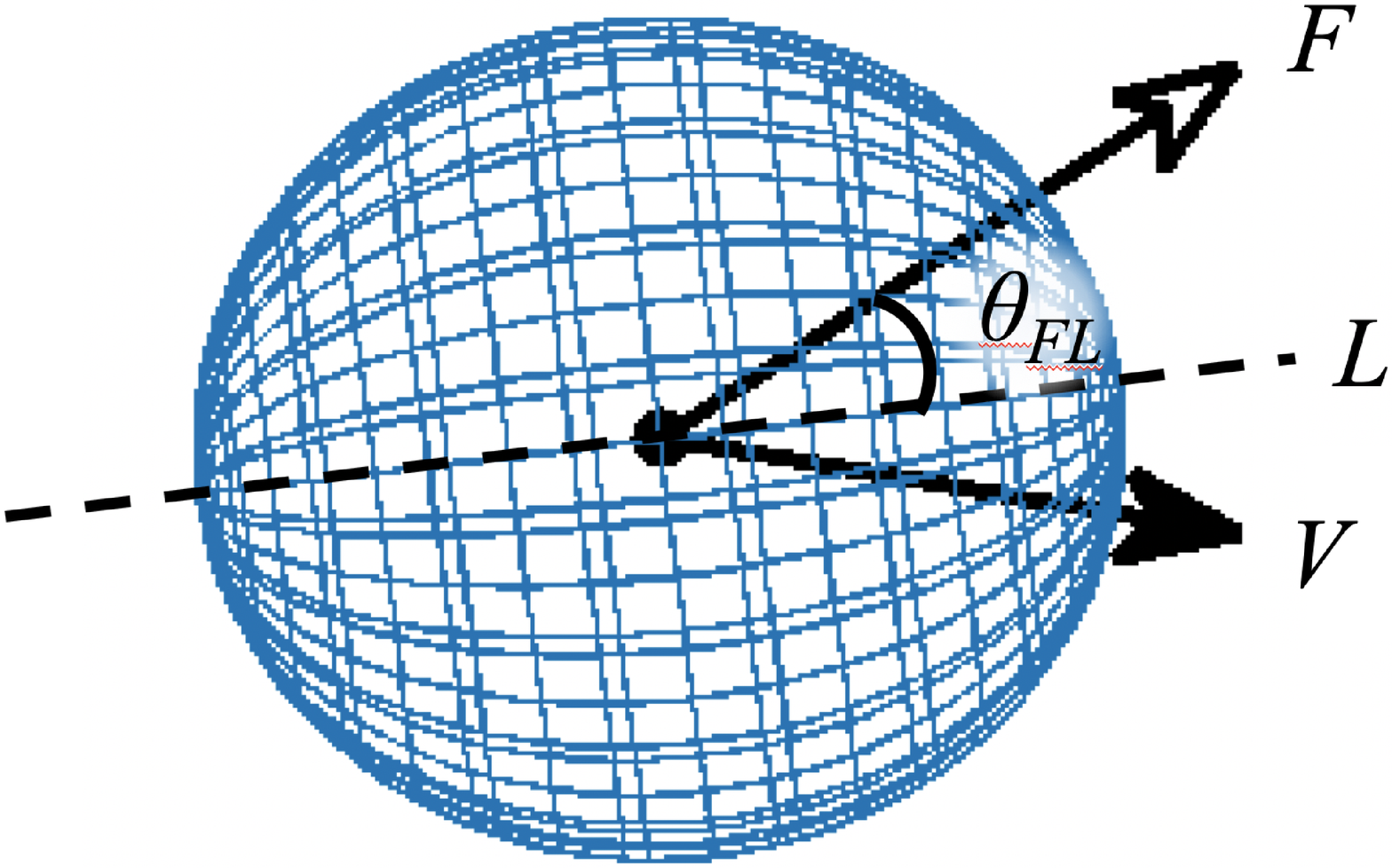}
\end{center}
\caption{A representative configuration of the ellipsoid  made by the eigenvalues and eigenvectors of the Eddington tensor.
The dashed line indicates the longest principal axis $L$.
The white and black arrow heads respectively indicate the directions of the neutrino energy flux $F$ and the matter velocity $V$.
The angle between $F$ and $L$ is $\theta_{FL}$.
\label{fig:ellipsoid}}
\end{figure}

The eigenvalues and eigenvectors of the Eddington tensor are calculated by the Jacobi method \citep[][]{Iwakami2020}.
In the principal-axis analysis, they are visualized as the ellipsoid:
\begin{equation}
\left(\frac{p^1}{\lambda^1}\right)^2 + \left(\frac{p^2}{\lambda^2}\right)^2 + \left(\frac{p^3}{\lambda^3}\right)^2 = 1,
\label{eq:ellipsoidorg}
\end{equation}
where $\lambda^j$ and $p^j$ are respectively the $j$-th eigenvalue and the momentum projected onto the $j$-th eigenvector.
The surface of the triaxial ellipsoid is parametrically expressed as
\begin{equation}
\left(\begin{array}{c}
      p^\theta\\
      p^\phi\\
      p^r
      \end{array}
\right)=
\left(\begin{array}{ccc}
      \mathcal{V}^{\theta 1} & \mathcal{V}^{\theta 2} & \mathcal{V}^{\theta 3} \\
      \mathcal{V}^{\phi 1} & \mathcal{V}^{\phi 2} & \mathcal{V}^{\phi 3}\\ 
      \mathcal{V}^{r 1} & \mathcal{V}^{r 2} & \mathcal{V}^{r 3}
\end{array}
\right)
\left(\begin{array}{l}
      \lambda^1 \cos a \cos b \\
      \lambda^2 \cos a \sin b \\
      \lambda^3 \sin a 
      \end{array}
\right), \label{eq:ellipsoid}
\end{equation}
where $a$ and $b$ are the parameters.
The components of the momentum vectors $(p^\theta, p^\phi, p^r)$ are the coordinates of a point on the ellipsoidal surface and $(\mathcal{V}^{\theta j}, \mathcal{V}^{\phi j}, \mathcal{V}^{rj})$ are the $\theta$-, $\phi$- and $r$- components of the $j$-th eigenvector.
We choose $\lambda^3$ as the largest of these eigenvalues to represent the longest axis of the ellipsoid.

Using Eq.~(\ref{eq:ellipsoid}), an ellipsoid can be drawn as shown in Figure \ref{fig:ellipsoid}.
The wireframe shape becomes a sphere, ellipsoid, and line in the optically thick limit, transition region, and optically thin limit, respectively.
In the figure, the longest principal axis is represented by $L$ whereas the neutrino energy flux and matter velocity are denoted by $F$ and $V$, respectively.
The angle between $F$ and $L$ is $\theta_\mathrm{FL}$.

\section{Principal-axis analysis in 1D \label{sec:1d}}

\begin{figure*}[ht!]
\begin{center}
\includegraphics[width=\hsize]{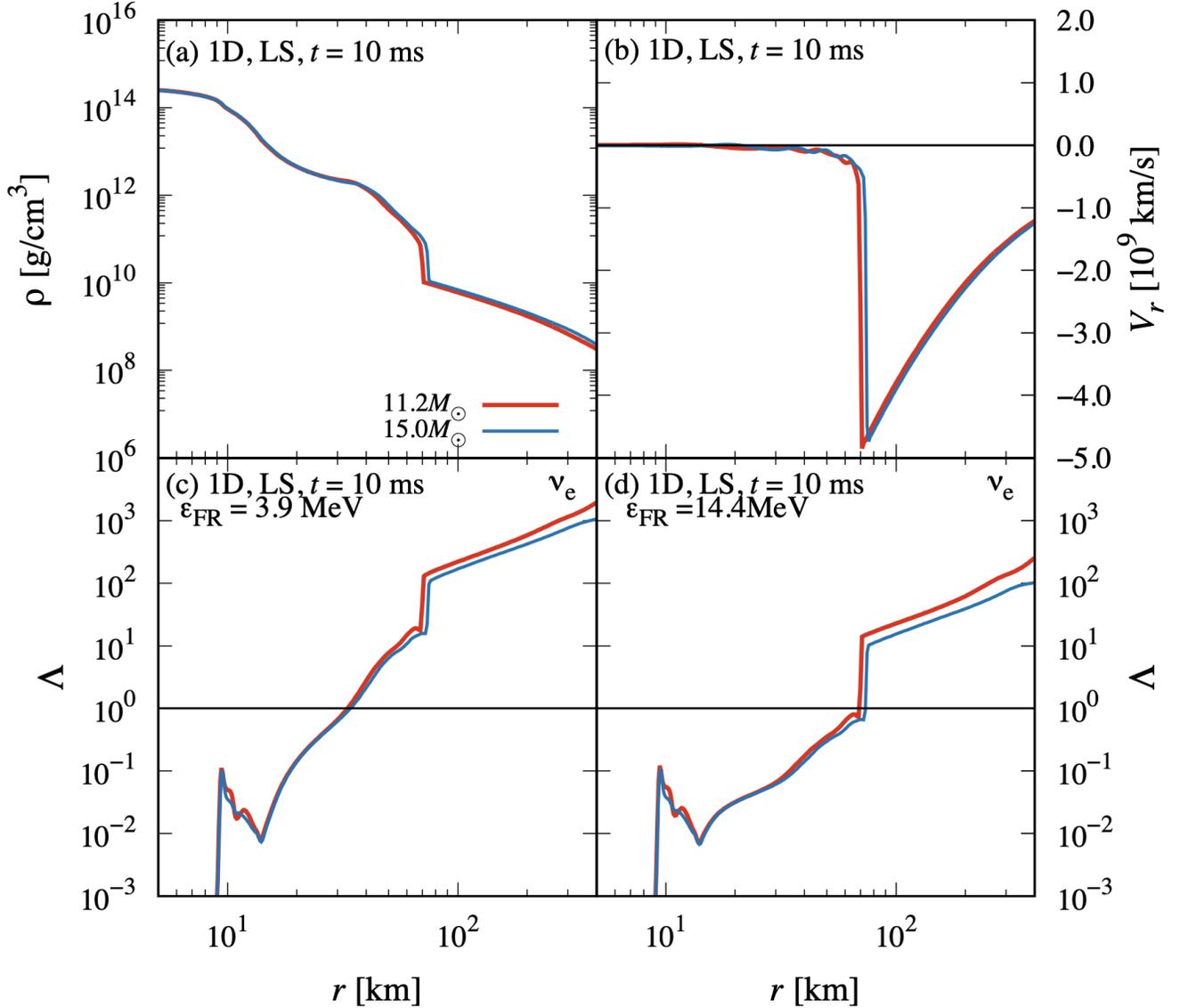}
\end{center}
\caption{Radial distributions of various variables for 11.2$M_\odot$ and 15.0$M_\odot$ progenitor models at $t=10$ ms, where $\rho$, $v_r$, $S$, $Y_L$, and $\Lambda$ are the density, radial velocity, entropy, lepton fraction, and mean free path divided by the radius, respectively.
\label{fig:hydro1d}}
\end{figure*}

\begin{figure*}[ht!]
\begin{center}
\includegraphics[width=\hsize]{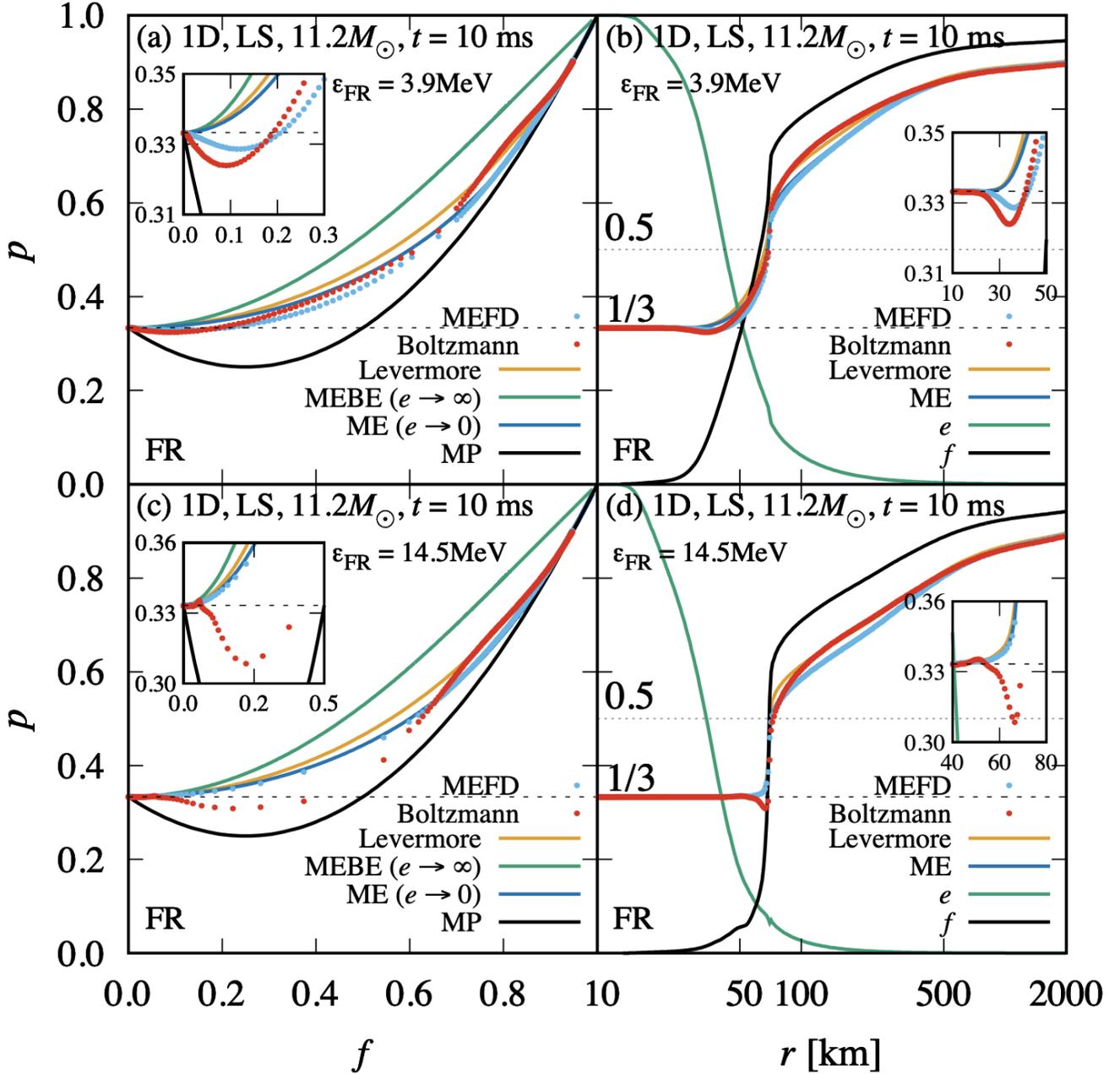}
\end{center}
\caption{The Eddington factor $p$ as a function of the flux factor $f$ (left panels) and the radial profiles of $p$ (right panels) for $\epsilon_{FR}$=3.9 (upper panels) and 14.5 (lower panels) MeV at $t=10$ ms  for $\nu_e$.
The red and light blue dots respectively indicate the Eddington factors obtained by Boltzmann transport and MEFD closure. In the left panels, the solid lines denote the closures of Levermore (yellow), MEBE in the limit of $e\rightarrow \infty$ (green), ME in the classical limit of $e\rightarrow 0$ (blue), and MP (black). In the right panels, the green and black lines respectively indicate the phase space occupancy $e$ and energy flux $f$.
\label{fig:pfr1d}}
\end{figure*}

\begin{figure*}[ht!]
\begin{center}
\includegraphics[width=\hsize]{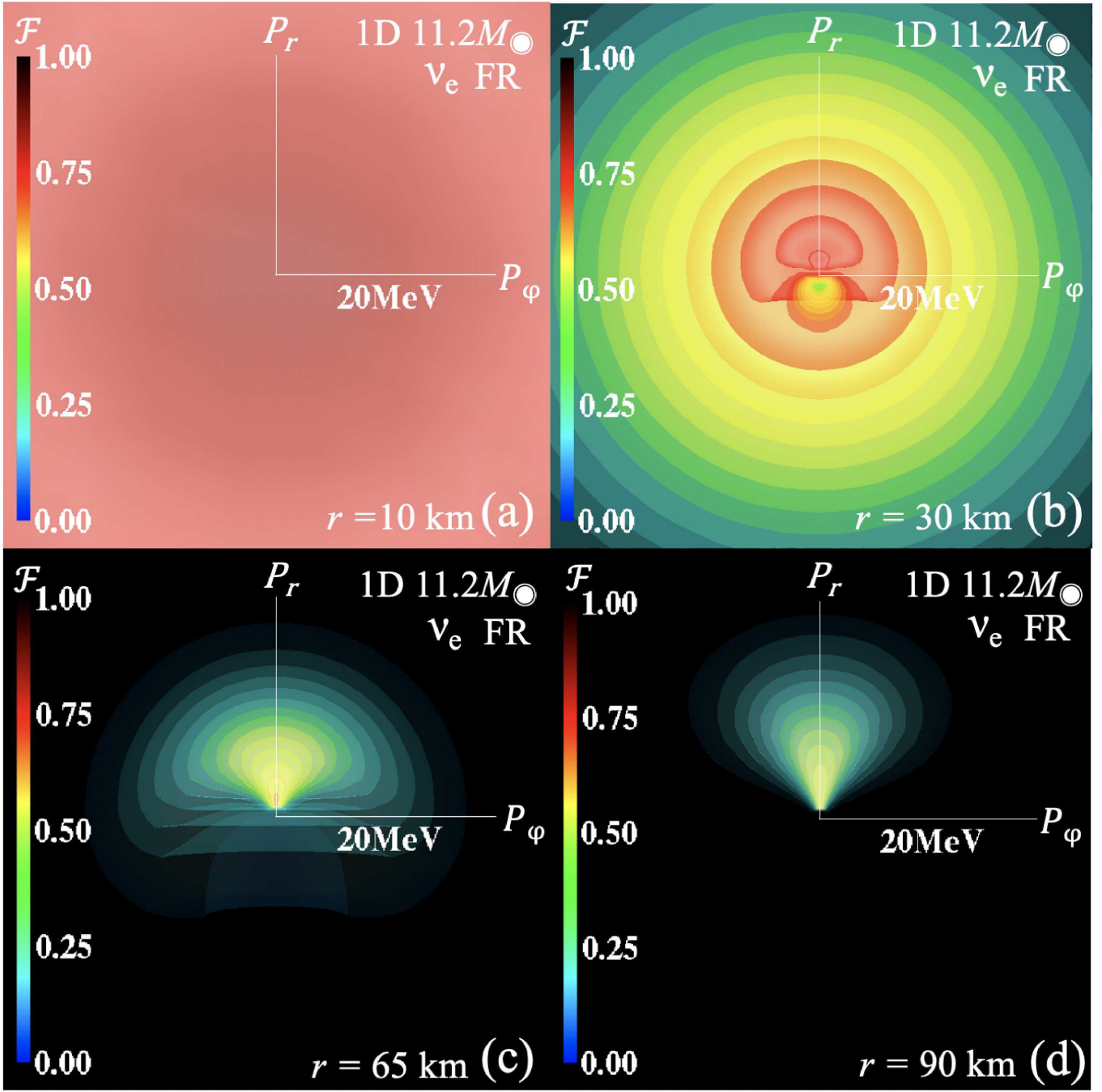}
\end{center}
\caption{The momentum space distribution of $\mathcal{F}$ in the FR for $\nu_e$ at $r=10$, 30, 65, and 90 km.
The distance from the origin is the neutrino energy $\epsilon_\mathrm{FR}$ and the lengths of the white axes correspond to 20 MeV.
\label{fig:fnue1d}}
\end{figure*}

\begin{figure*}[ht!]
\begin{center}
\includegraphics[width=\hsize]{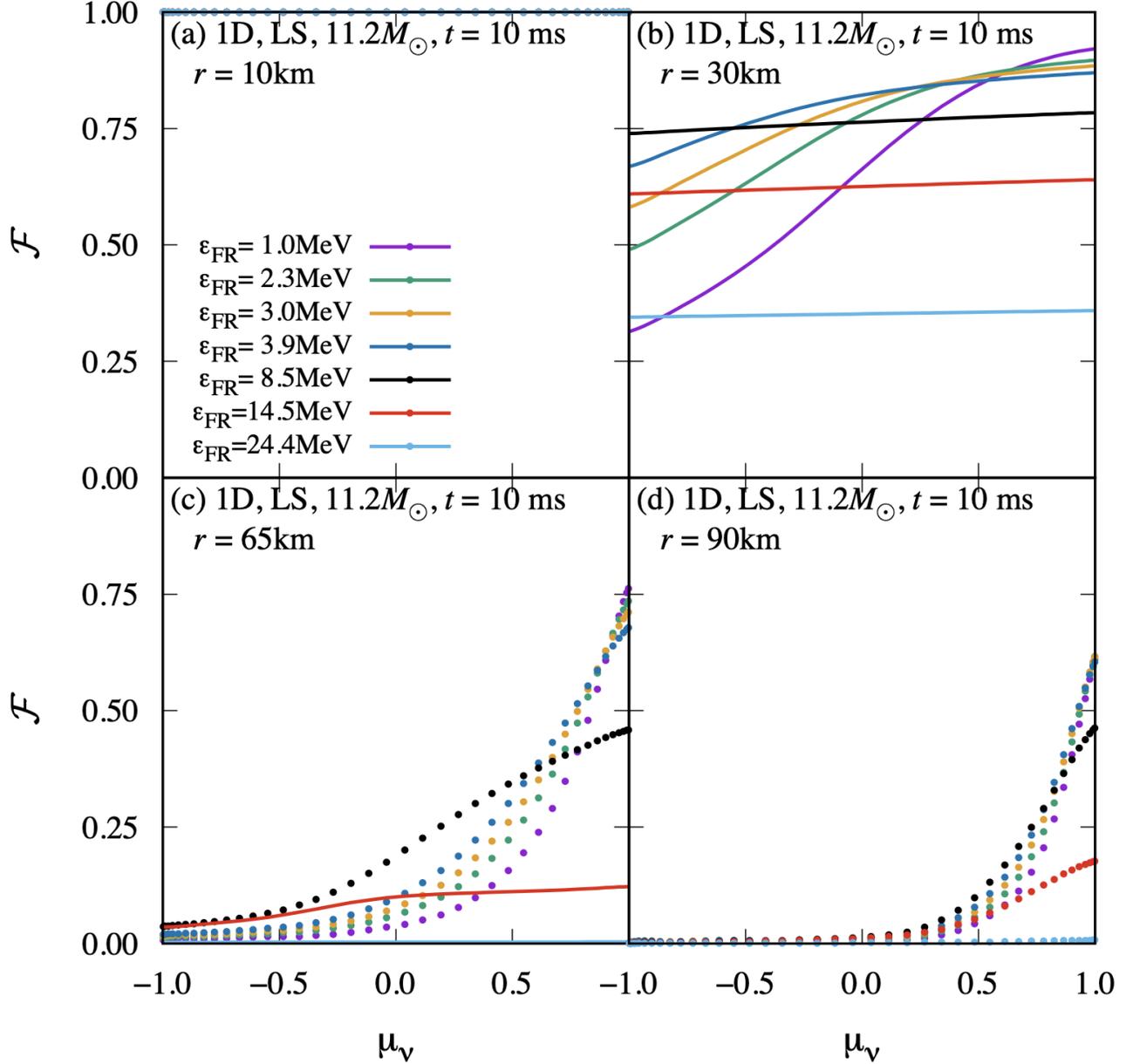}
\end{center}
\caption{The angular distribution of $\mathcal{F}$ for $\nu_e$ at $r=10$, 30, 65, and 90 km. The symbol $\mu_\nu(=\cos\theta_\nu)$ is the cosine of the polar angle in momentum space.
The solid lines and dots are the angular distributions corresponding to $p<1/3$ and $p\ge1/3$, respectively.
The specific values of $p$ are 0.327 (purple), 0.322 (green), 0.324 (yellow), 0.327 (blue), 0.331 (black), 0.333 (red), 0.333 (light blue) at $r=30$ km (b), and 0.311 (red), 0.331 (light blue) at $r=65$ km (c).
\label{fig:fnueang}}
\end{figure*}

\begin{figure*}[ht!]
\begin{center}
\includegraphics[width=\hsize]{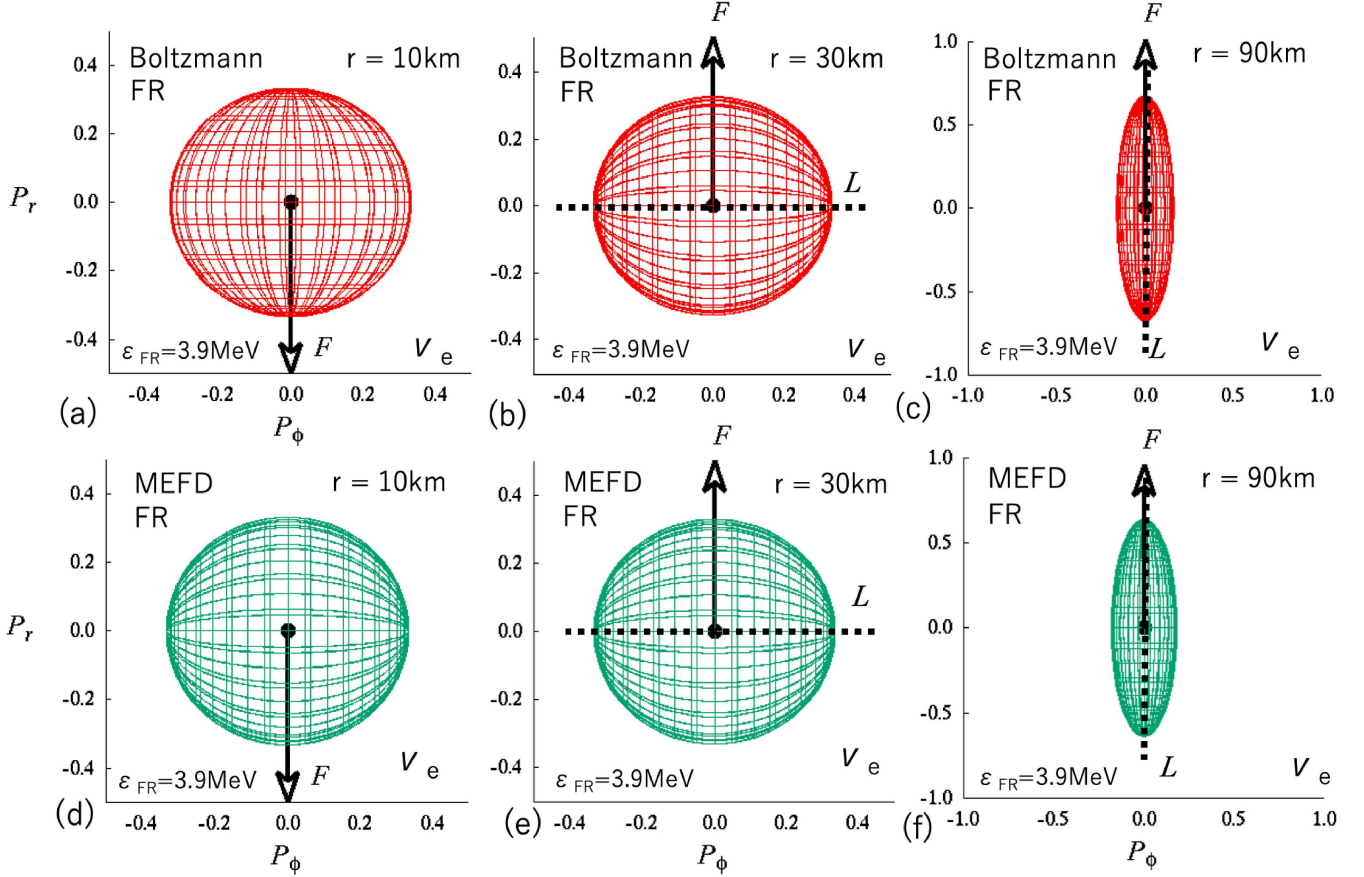}
\end{center}
\caption{The ellipsoid of the Eddington tensors obtained in the Boltzmann simulation (top panels) and those evaluated in the MEFD closure approximation (lower panels) for $\epsilon_\mathrm{FR}$=3.9 MeV  at $r=$10 km (left panels), 30 km (middle panels), and 90 km (right panels).
\label{fig:ellipsoid1Den004}}
\end{figure*}

\begin{figure*}[ht!]
\begin{center}
\includegraphics[width=\hsize]{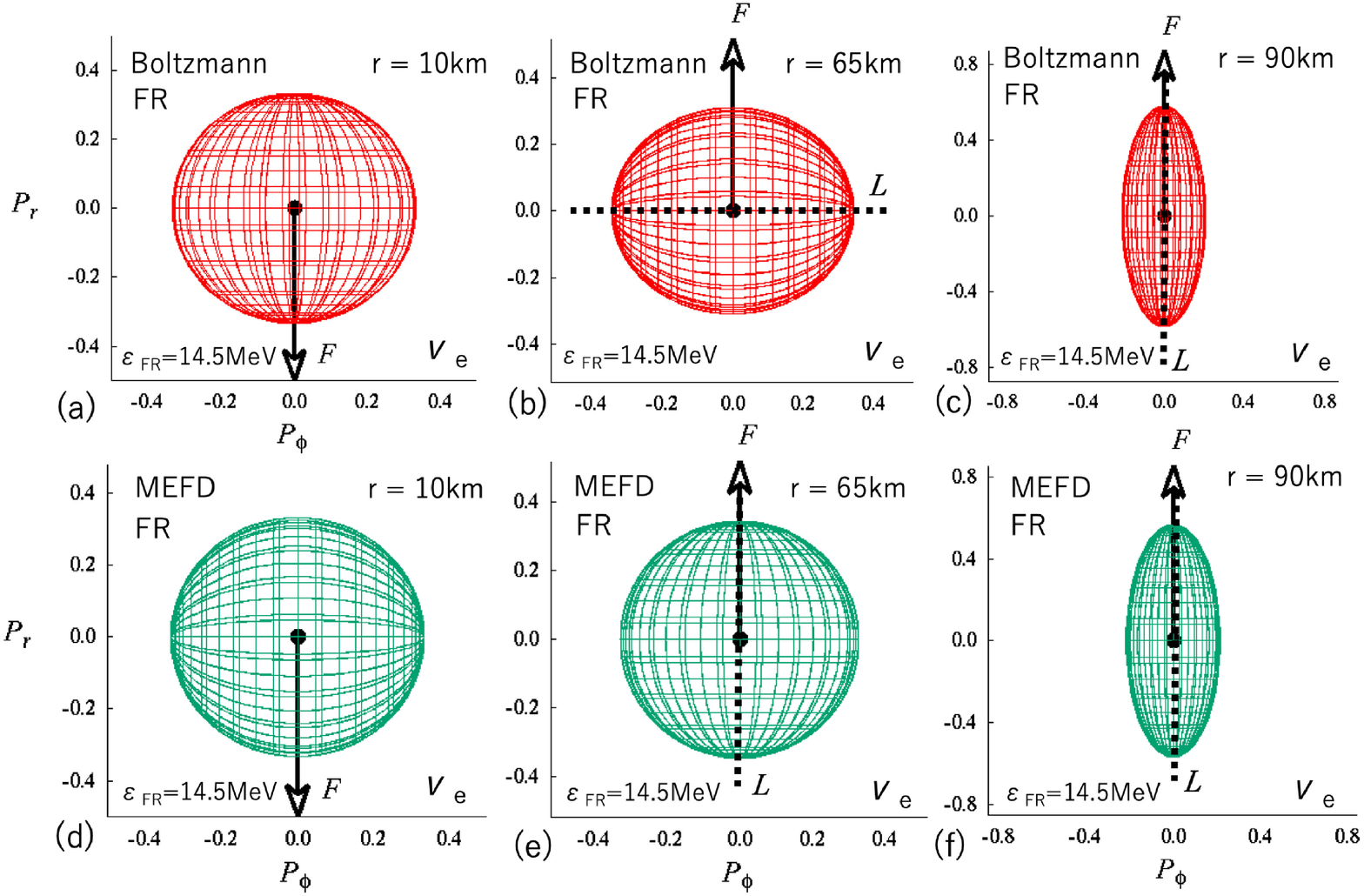}
\end{center}
\caption{As in the previous figure but for the neutrino energy $\epsilon_\mathrm{FR}$=14.5 MeV with the middle panels showing radius $r=65$ km.
\label{fig:ellipsoid1Den009}}
\end{figure*}

This section is devoted to the screening of the closure relations commonly used in the literature and listed in section \ref{sec:closures} by using the results of 1D Boltzmann simulations.
Such comparisons have been made over the years by other authors in their own contexts \citep[e.g.][]{Janka1992, Smit2000, Murchikova2017} and the results obtained here are consistent with the previous ones.
We also apply the principal axis analysis to the MEFD closure, which is singled out to be the best and will be adopted also in the 2D analysis later.
A particular attention is paid to the shape change of the ellipsoid (Eq.~(\ref{eq:ellipsoidorg})) in the MEFD closure, which is shared in 2D and hence will be useful later.

Figure \ref{fig:hydro1d} shows the radial profiles of the various quantities for the 11.2$M_\odot$ and 15.0$M_\odot$ models at 10ms post bounce, where $S$, $Y_L$, and $\Lambda$ are the entropy, lepton fraction, and mean free path divided by the radius, respectively.
The shock wave is located around $r=70$ km.
Since the profiles are not much different from each other for $r \lesssim 100$ km, except for the entropy above the shock wave, only the results for the 11.2$M_\odot$ model are presented in this section.

The Eddington factor $p$ for $\nu_e$ are shown as functions of $f$ (left panel) and $r$ (right panel) in Figure \ref{fig:pfr1d}.
The electron-neutrino energy $\epsilon_\mathrm{FR}$ is 3.9 MeV in the upper panels and 14.5 MeV in the bottom panels.
The latter is almost the average energy of $\nu_e$ just behind the shock wave at this time.
The Eddington factors $p_\mathrm{Boltzmann}$
and $p_\mathrm{MEFD}$ are plotted with red and light blue dots, respectively. 
In the left panels, the Levermore, MEBE ($e\rightarrow \infty$) and ME ($e\rightarrow 0$) closures are shown with colored solid lines, and the MP is indicated by a black line.
In the right panels, the green and black lines denote the phase space occupancy $e$ and flux factor $f$, respectively.

As pointed out in our previous paper \citep{Iwakami2020}, the regions of $p_\mathrm{Boltzmann}<1/3$ appear at $\Lambda \lesssim 1$ (Fig.~\ref{fig:hydro1d}(c) and (d)) for both 3.9 and 14.5 MeV (Fig.~\ref{fig:pfr1d}).
For $\epsilon_{FR}=3.9$ MeV, this region is overlapped with the region of $p_\mathrm{MEFD}<1/3$, which emerges inside the proto-neutron star (PNS) at $10<r<40$ km (Fig.~\ref{fig:pfr1d}(b)), where the flux factor is $0 \lesssim f \lesssim 0.2$ (Fig.~\ref{fig:pfr1d}(a)).
Note that $p_\mathrm{Levermore}$, $p_\mathrm{ME}$, and $p_\mathrm{MEBE}$ increase monotonically with increasing $f$ (Eqs.~(\ref{eq:p_ME}-\ref{eq:p_Levermore})) and are qualitatively different from the result of Boltzmann transport simulation.
The similar results were reported by \citet{Janka1992} and \citet{Smit2000}.
By contrast, at $\epsilon_{FR}=14.5$ MeV, the region of $p_\mathrm{Boltzmann}<1/3$ is wider, extending from inside the PNS up to behind the shock wave, where the range of the flux factor is $0.05 \lesssim f \lesssim 0.4$ (Fig.~\ref{fig:pfr1d}(c)).
Note that $p_\mathrm{MEFD}>1/3$ at 50 km $\lesssim r \lesssim$ 70 km (Fig.~\ref{fig:pfr1d}(d)). 
This is because $e < 0.5$ this time.
As a matter of fact, $p_\mathrm{MEFD}$ is known to be a monotonically increasing function of $f$ when $e < 0.5$ (Eq.~(\ref{eq:p_MEFD})).
Hence even the MEFD closure gives qualitatively wrong Eddington factors in such situations.

The Levermore closure seems to be best for $f\gtrsim0.7$ (Figs.~\ref{fig:pfr1d}(c) and (d)). 
As the Boltzmann transport simulation tends to be underresolved in momentum space (see Appendix~\ref{sec:resolution}), however, a closure that gives a more forward-peaked distribution is desirable in the optically thin region.
As was concluded also by \citet{Murchikova2017}, it is confirmed that no single closure can give an accurate estimate to the Eddington factor in the CCSN context.

Figure \ref{fig:fnue1d} shows the distribution $\mathcal{F}$ in momentum space for $\nue$ in the FR at $r=10$, 30, 65, and 90 km.
In this space, the distance from the origin, the angle from the $P_r$ axis, and the angle from the $P_\theta$ axis on the $P_\theta - P_\phi$ plane correspond to $\epsilon$, $\theta_\nu$, and $\phi_\nu$, respectively.
In Figure \ref{fig:fnueang}, we present $\mathcal{F}$ as a function of $\mu_\nu$ for $\nu_e$ where $\mu_\nu\  =\cos\theta_\nu$.
The neutrinos are uniformly distributed over the entire solid angle for all energy bins at $r=10$ km in the optically thick region (Figs.~\ref{fig:fnue1d}(a) and \ref{fig:fnueang}(a)).
As the radius increases, more forward-peaked distributions start to emerge in the lower energy neutrinos (Figs.~\ref{fig:fnue1d}(b), (c), and (d)).
The hemispheric distribution, in which $\mathcal{F}$ is rather uniformly distributed in the forward and not so much distributed in the backward directions, can be observed in the blue line for $\epsilon_\mathrm{FR} = 3.9$ MeV at $r=30$ km (Fig.~\ref{fig:fnueang}(b)) and in the red line for $\epsilon_\mathrm{FR} = 14.5$ MeV at $r=65$ km (Fig.~\ref{fig:fnueang}(c)).
Although the Eddington factor tends to be smaller than 1/3 when the angular distribution is hemispheric, the almost uniform distributions denoted by the black, red, and light blue lines at $r=30$ km (Fig.~\ref{fig:fnueang}(b)) and by the light blue lines at $r=65$ km (Fig.~\ref{fig:fnueang}(c)) also give $p_\mathrm{Boltzmann} \lesssim 1/3$.
What is essential to produce $p < 1/3$ is that neutrinos are abundant at $\mu_{\nu} \approx 0$, since they contribute to the zeroth moment but not to the second moment, rendering the Eddington tensor smaller.

Figures \ref{fig:ellipsoid1Den004} and \ref{fig:ellipsoid1Den009} present the ellipsoids of the Eddington tensors obtained in the Boltzmann simulation (top panels) and those given by the MEFD closure (lower panels) at three radii for $\epsilon_\mathrm{FR}=3.9$ and 14.5 MeV, respectively.
The Eddington tensors are evaluated in the FR.
In the left panels (Figs.~\ref{fig:ellipsoid1Den004}(a), \ref{fig:ellipsoid1Den004}(d), \ref{fig:ellipsoid1Den009}(a) and \ref{fig:ellipsoid1Den009}(d)), on the other hand, an almost-complete sphere is obtained for both Boltzmann and MEFD at $r=10$ km, where $\Lambda \lesssim 0.1$ (Figs.~\ref{fig:hydro1d}(e) and (f)).~\footnote{Since the accuracy of the Jacobi method to obtain eigenvalues and eigenvectors is $O(10^{-8})$ in the principal-axis analysis, the direction of $L$, the longest principal axis, is rather meaningless at $r\lesssim 10$ km.}
In the right panels (Figs.~\ref{fig:ellipsoid1Den004}(c), \ref{fig:ellipsoid1Den004}(f), \ref{fig:ellipsoid1Den009}(c) and \ref{fig:ellipsoid1Den009}(f)), the vertically elongated ellipsoid, in which the longest axis $L$ is parallel to $F$, is drawn for both Boltzmann and MEFD at $r=90$ km, where $\Lambda \gtrsim 10$ (Figs.~\ref{fig:hydro1d}(e) and (f)).
In the middle panels, the horizontally elongated ellipsoid, where $L$ is perpendicular to $F$, is derived for both Boltzmann and MEFD at $\epsilon_\mathrm{FR}=3.9$ MeV (Figs.~\ref{fig:ellipsoid1Den004}(b) and (e)) and for Boltzmann alone at $\epsilon_\mathrm{FR}=14.5$ MeV (Fig.~\ref{fig:ellipsoid1Den009}(b)).
At this radius $\Lambda \approx 1$ (Figs.~\ref{fig:hydro1d}(e) and (f)) and matter is neither optically thick nor thin.
The difference between Boltzmann and MEFD for average-energy neutrinos is a confirmation that the longest principal axis $L$ is perpendicular to $F$ if the flux factor $p$ is smaller than 1/3 and the MEFD closure fails to reproduce it.

\section{Principal-axis analysis in 2D\label{sec:2d}}

\begin{figure*}[ht!]
\begin{center}
\includegraphics[width=\hsize]{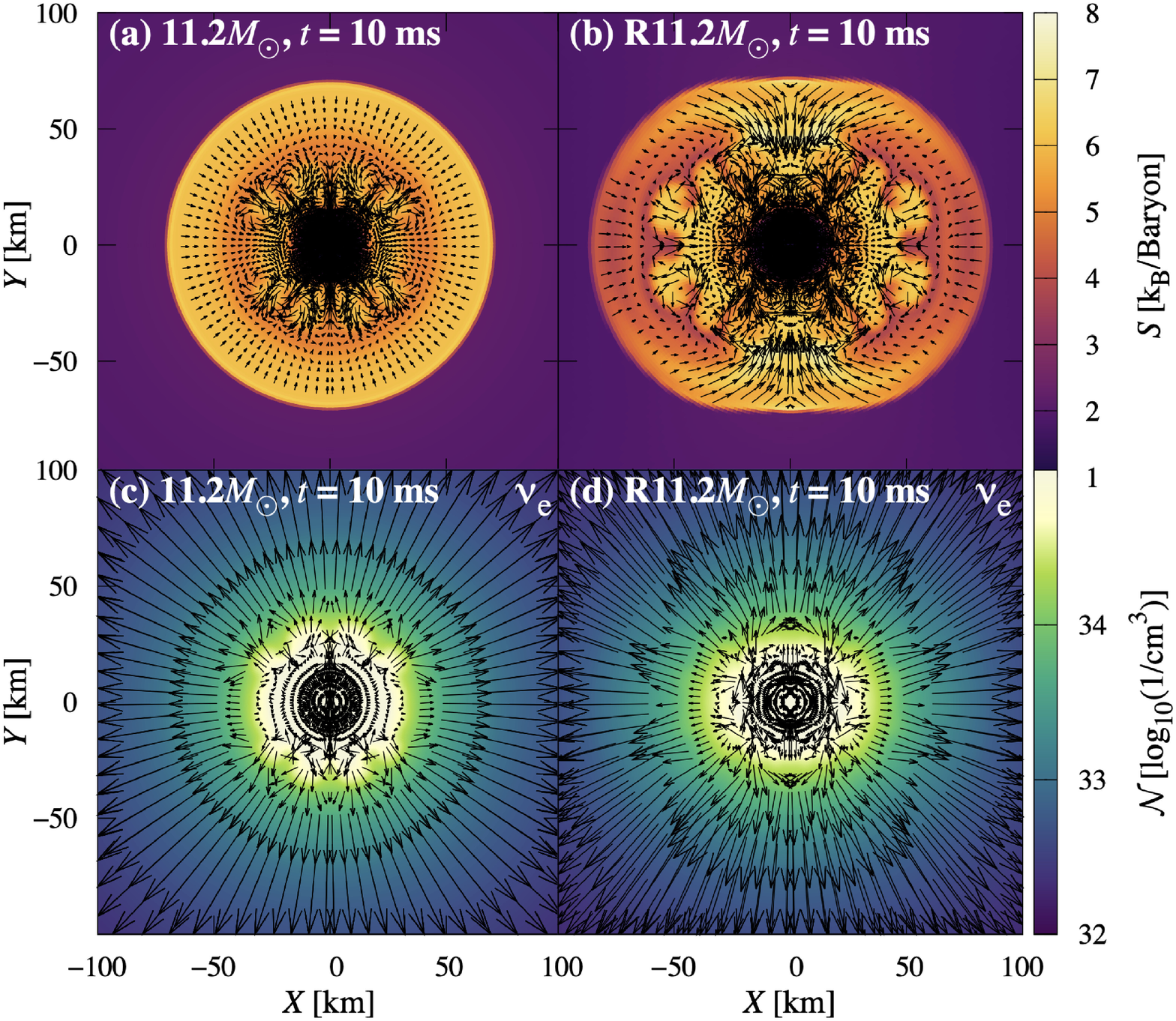}
\end{center}
\caption{Color maps of entropy $S$ with the fluid velocity vectors $V$ (upper panels) and the number density $\mathcal{N}$ of $\nu_\mathrm{e}$ with its average velocity vectors $V_{\nu}$ (lower panels) for non-rotating (left panels) and rotating (right panels) 11.2$M_\odot$ models at $t=10$ ms.
\label{fig:EVNF2D11_2M10ms}}
\end{figure*}

\begin{figure*}[ht!]
\begin{center}
\includegraphics[width=\hsize]{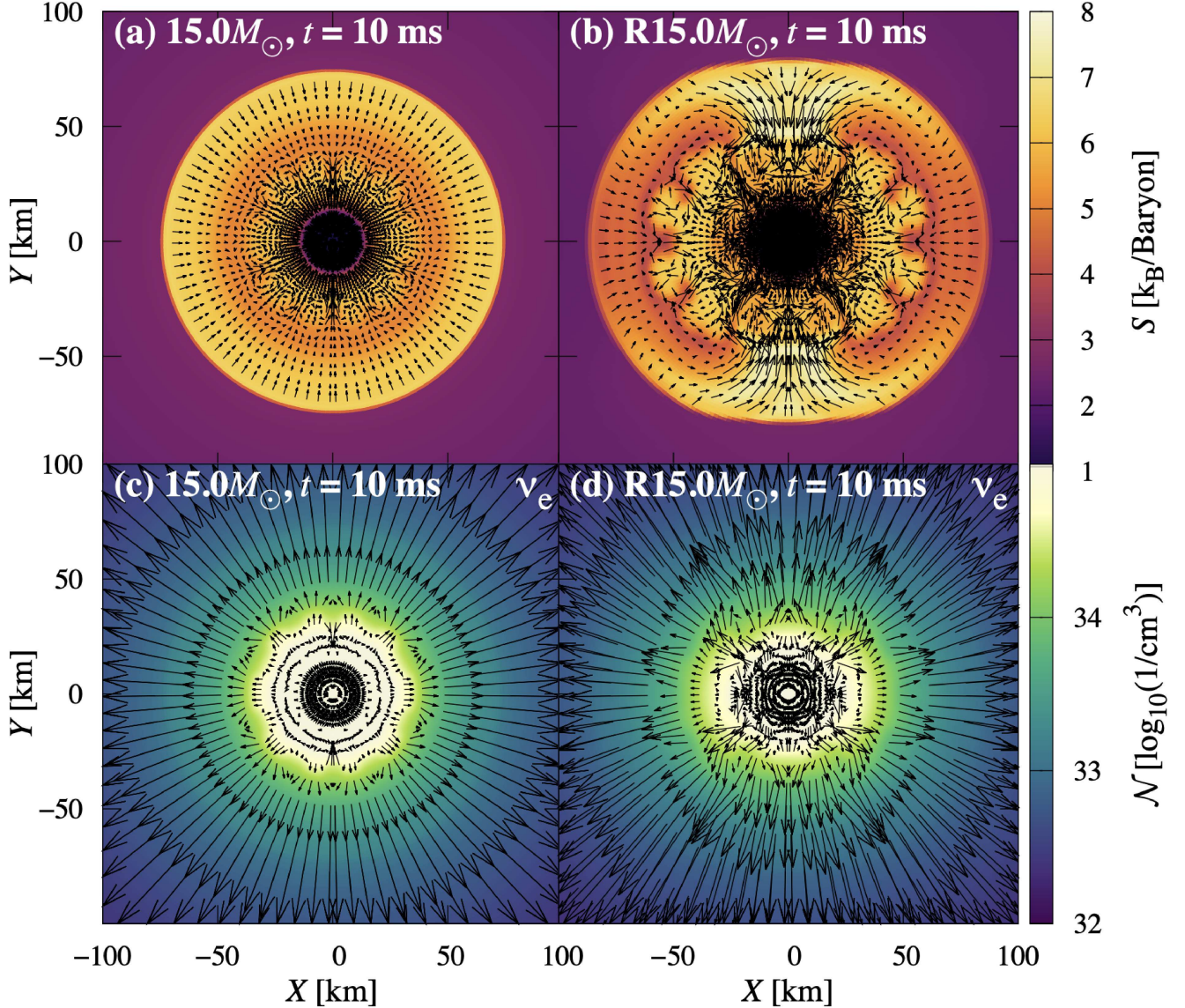}
\end{center}
\caption{As in Figure \ref{fig:EVNF2D11_2M10ms} but for the progenitor mass 15.0$M_\odot$.
\label{fig:EVNF2D15_0M10ms}}
\end{figure*}

\begin{figure*}[ht!]
\begin{center}
\includegraphics[width=\hsize]{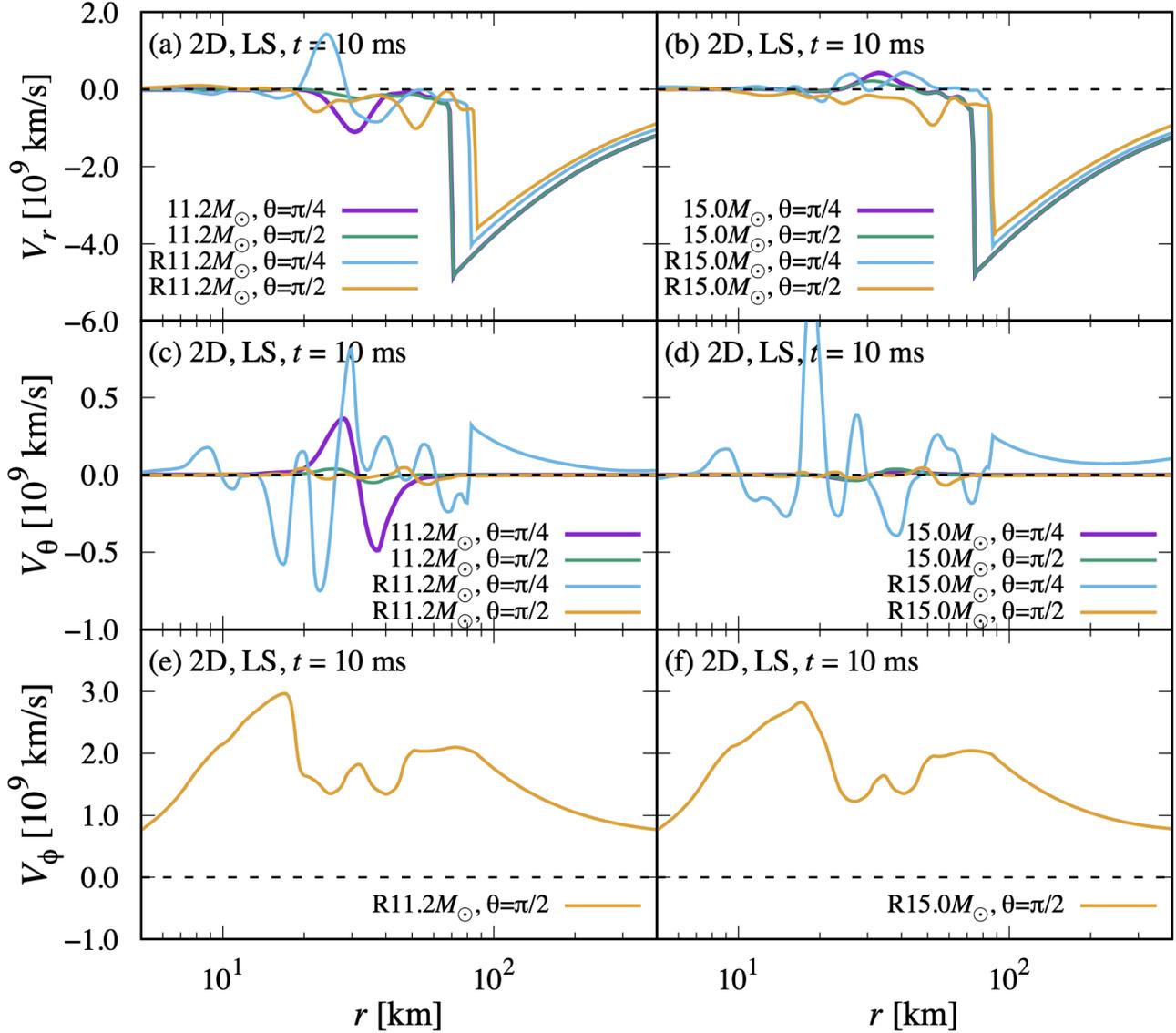}
\end{center}
\caption{Radial distributions of velocity components for 11.2$M_\odot$ and 15.0$M_\odot$ progenitor models at $t=10$ ms, where $V_r$, $V_\theta$, and $V_\phi$ are the radial, polar, and azimuthal matter velocities, respectively.
\label{fig:hydrovel2d10ms}}
\end{figure*}

\begin{figure*}[ht!]
\begin{center}
\includegraphics[width=\hsize]{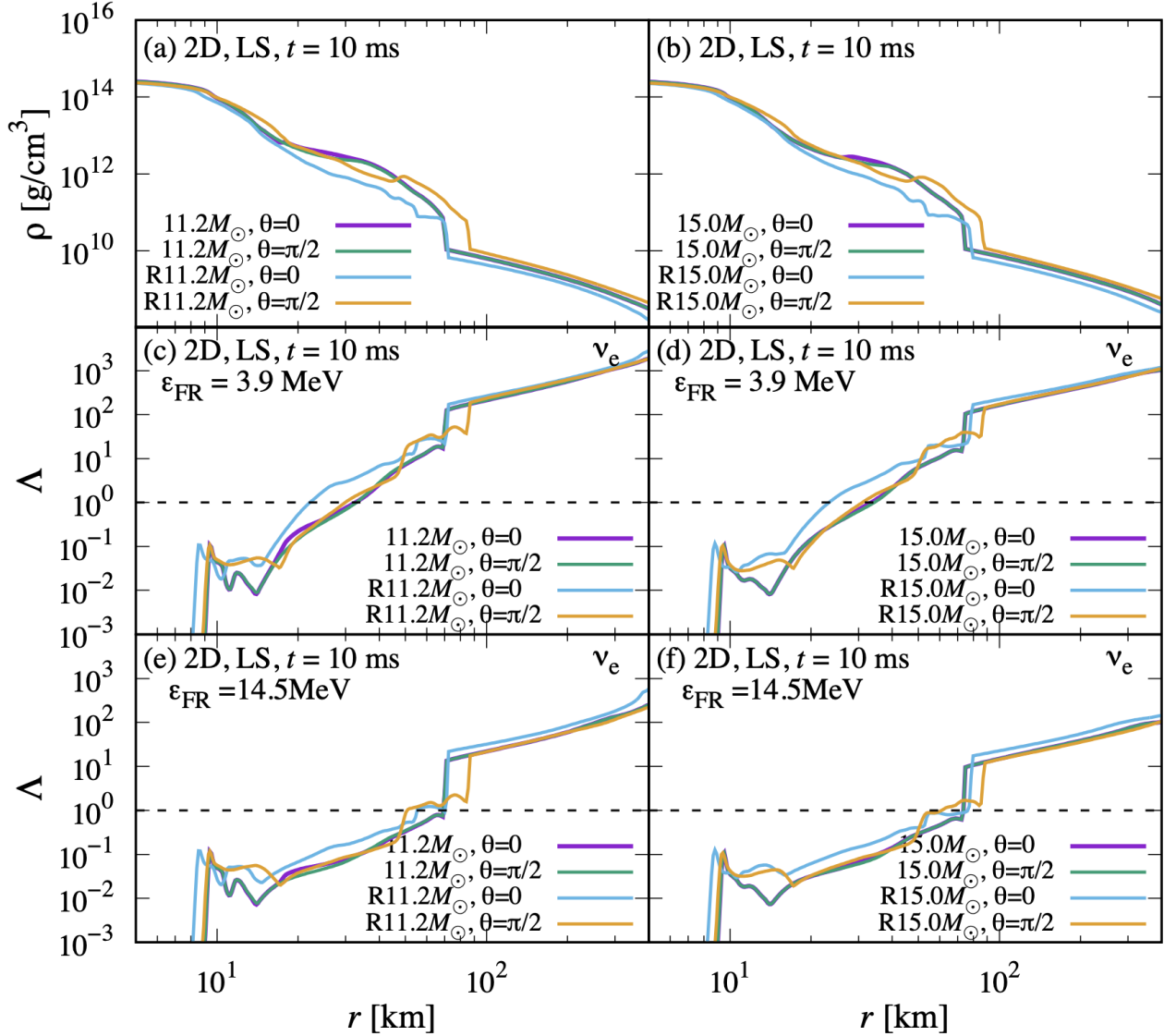}
\end{center}
\caption{Radial distributions of the density $\rho$ and the mean free path divided by the radius $\Lambda$ for 11.2$M_\odot$ and 15.0$M_\odot$ progenitor models at $t=10$ ms.
\label{fig:hydrorho2d10ms}}
\end{figure*}

\begin{figure*}[ht!]
\begin{center}
\includegraphics[width=\hsize]{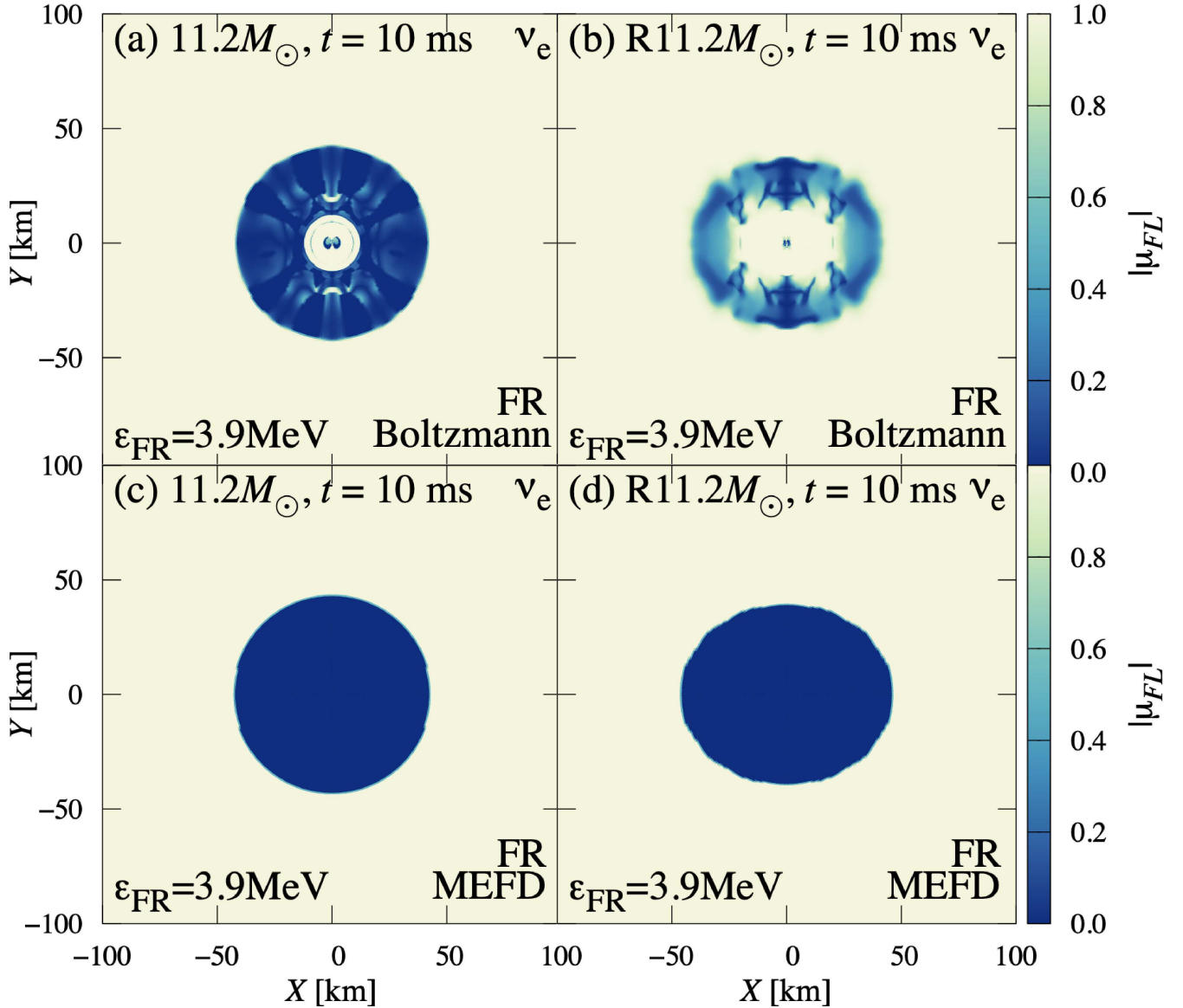}
\end{center}
\caption{Color maps of $|\mu_{FL}|$ in the FR for $\epsilon_\mathrm{FR}=3.9$ MeV at $t=10$ ms for the non-rotating (left panels) and rotating (right panels) 11.2$M_\odot$ models,  where $\mu_{FL}=\cos\theta_{FL}$.
The results obtained from the Boltzmann transport and MEFD closure are shown in the upper and lower panels, respectively.
\label{fig:INFR2DLS11_2en004}}
\end{figure*}

\begin{figure*}[ht!]
\begin{center}
\includegraphics[width=\hsize]{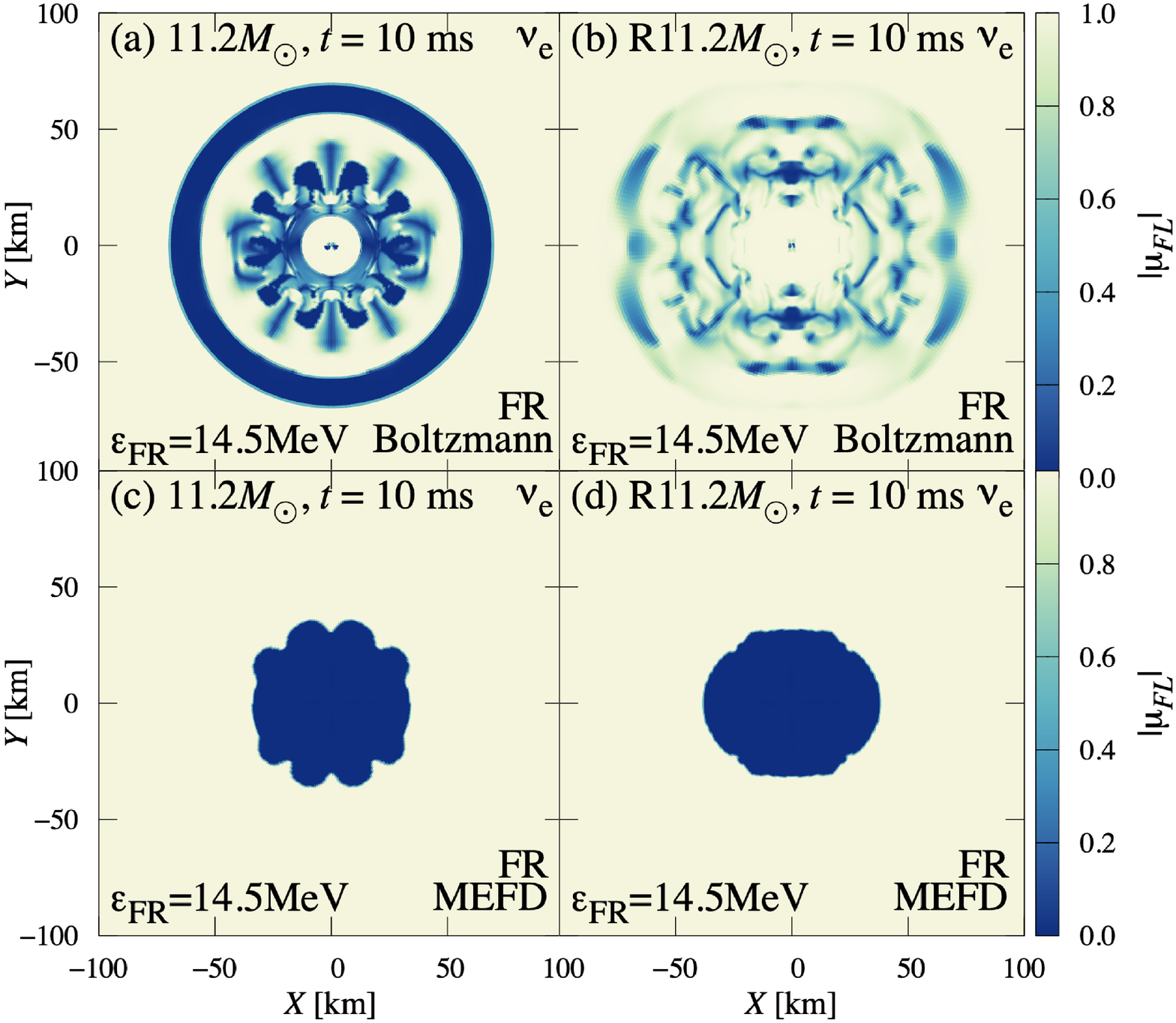}
\end{center}
\caption{As in Figure \ref{fig:INFR2DLS11_2en004} but for the neutrino energy $\epsilon_\mathrm{FR}=14.5$ MeV.
\label{fig:INFR2DLS11_2en009}}
\end{figure*}

\begin{figure*}[ht!]
\begin{center}
\includegraphics[width=\hsize]{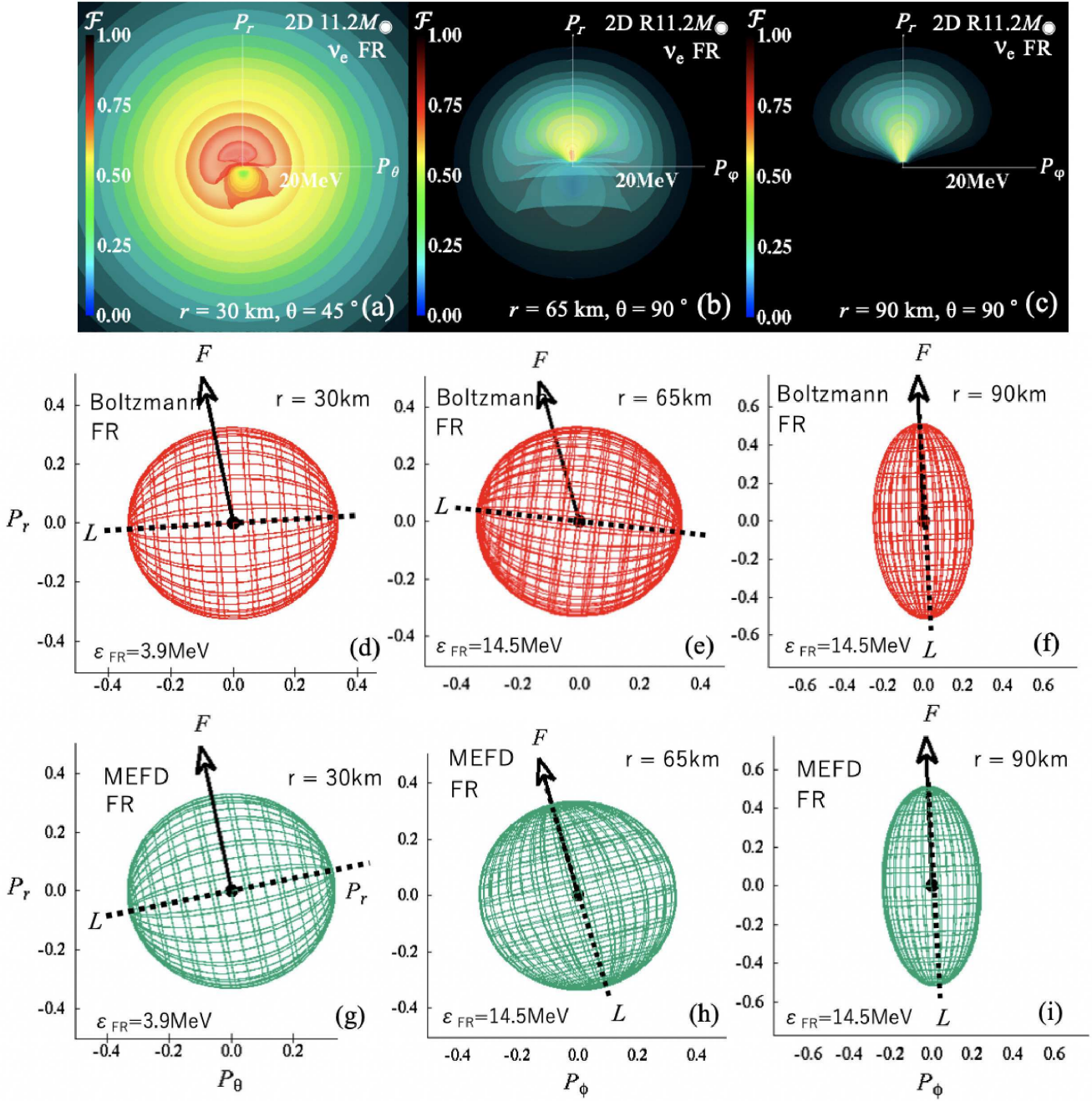}
\end{center}
\caption{Distribution functions $\mathcal{F}$ of $\nu_e$ in momentum space (upper panels), and the corresponding ellipsoids of the Eddington tensors obtained in the Boltzmann transport (middle panels) and those evaluated in the MEFD closure (lower panels) for the non-rotating model at $r=30$ km (left panels) and for the rotating model at $r=65$ km (middle panels) and 90 km (right panels).
The neutrino energy $\epsilon_\mathrm{FR}$ is 3.9 MeV for the non-rotating model and 14.5 MeV for the rotating model.
The reference frame is the FR.
\label{fig:NeutrinoEllipsoidFR}}
\end{figure*}

\begin{figure*}[ht!]
\begin{center}
\includegraphics[width=\hsize]{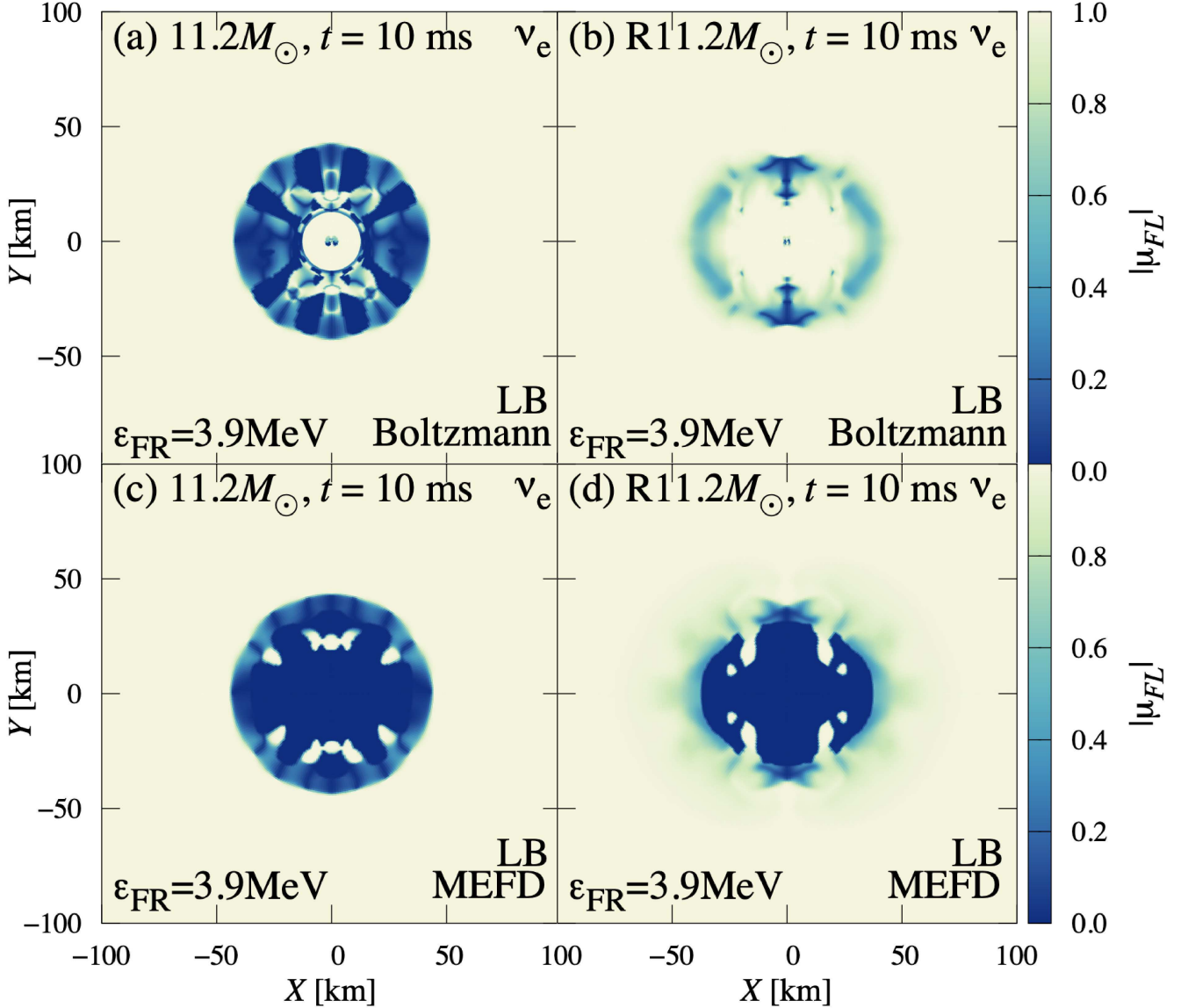}
\end{center}
\caption{Color maps of $|\mu_{FL}|$ in the LB for $\epsilon_\mathrm{FR}=3.9$ MeV at $t=10$ ms for the non-rotating (left panels) and rotating (right panels) 11.2$M_\odot$ models. The upper and lower panels show the results for Boltzmann and MEFD, respectively. 
\label{fig:INLB2DLS11_2en004}}
\end{figure*}

\begin{figure*}[ht!]
\begin{center}
\includegraphics[width=\hsize]{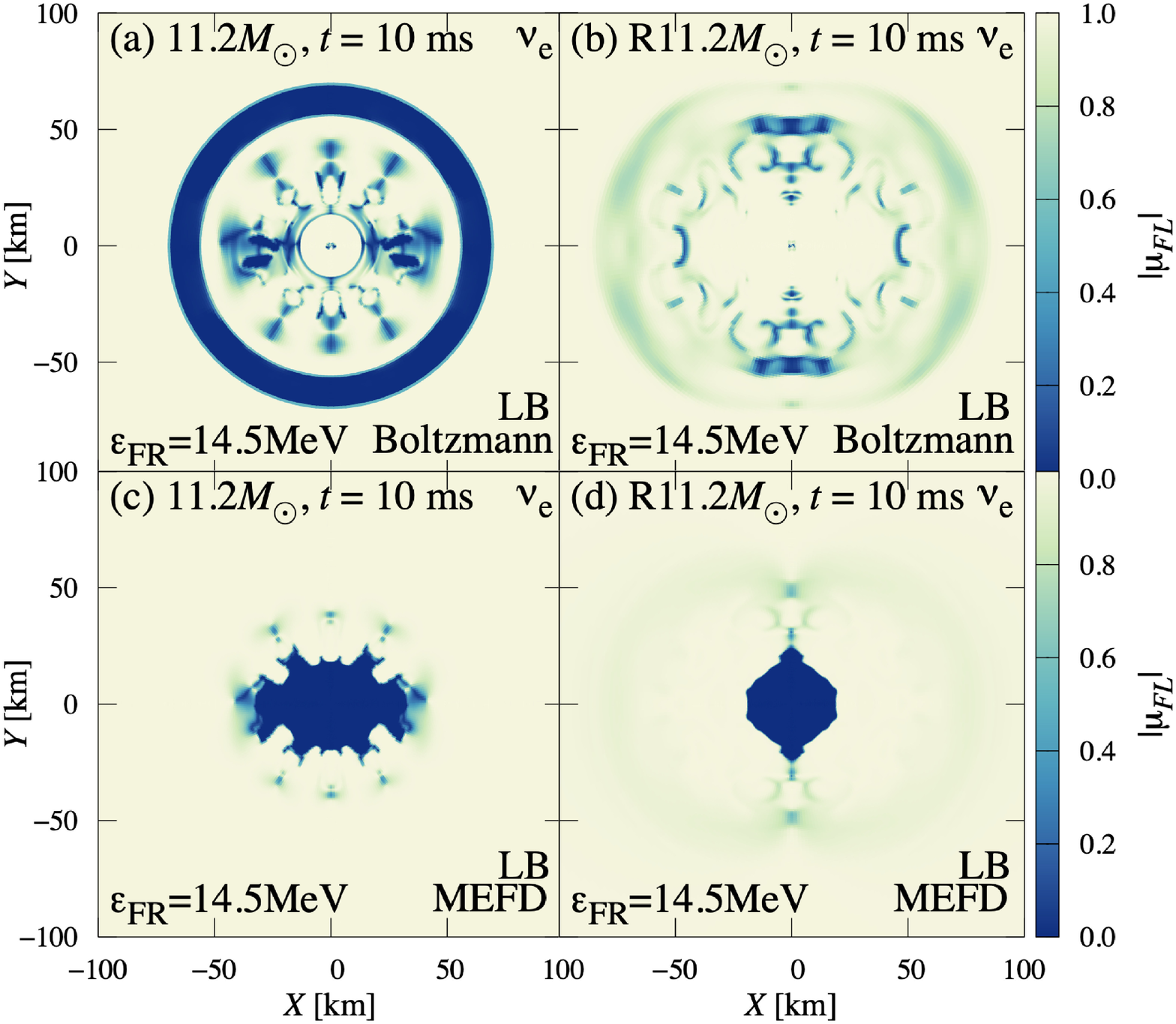}
\end{center}
\caption{As in Figure \ref{fig:INLB2DLS11_2en004} but for the neutrino energy $\epsilon_\mathrm{FR}=14.5$ MeV.
\label{fig:INLB2DLS11_2en009}}
\end{figure*}

\begin{figure*}[ht!]
\begin{center}
\includegraphics[width=\hsize]{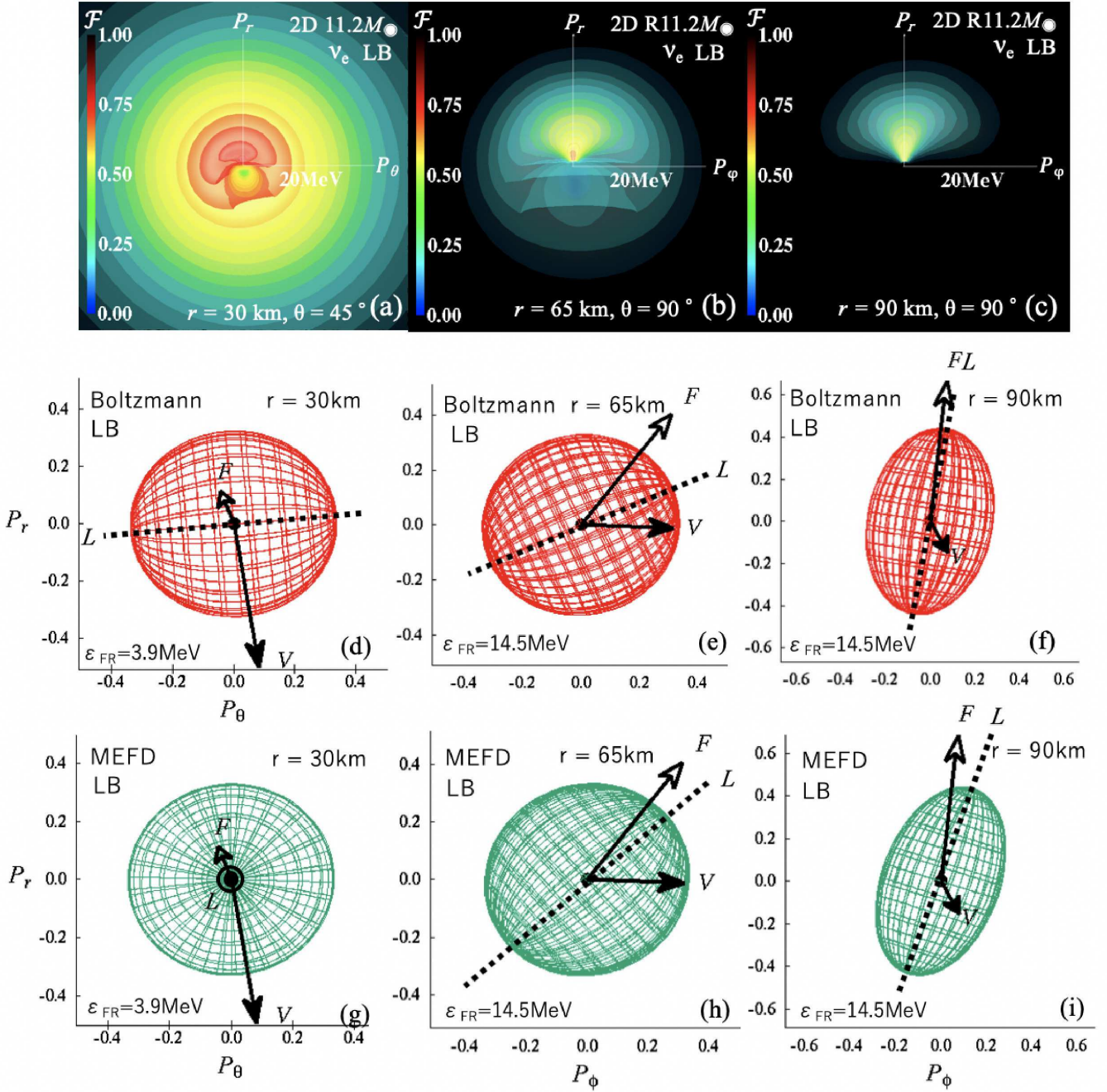}
\end{center}
\caption{As in Figure~\ref{fig:NeutrinoEllipsoidFR} but with the momentum space grids and Eddington tensors calculated in the LB.
\label{fig:NeutrinoEllipsoidLB}}
\end{figure*}

\begin{figure*}[ht!]
\begin{center}
\includegraphics[width=\hsize]{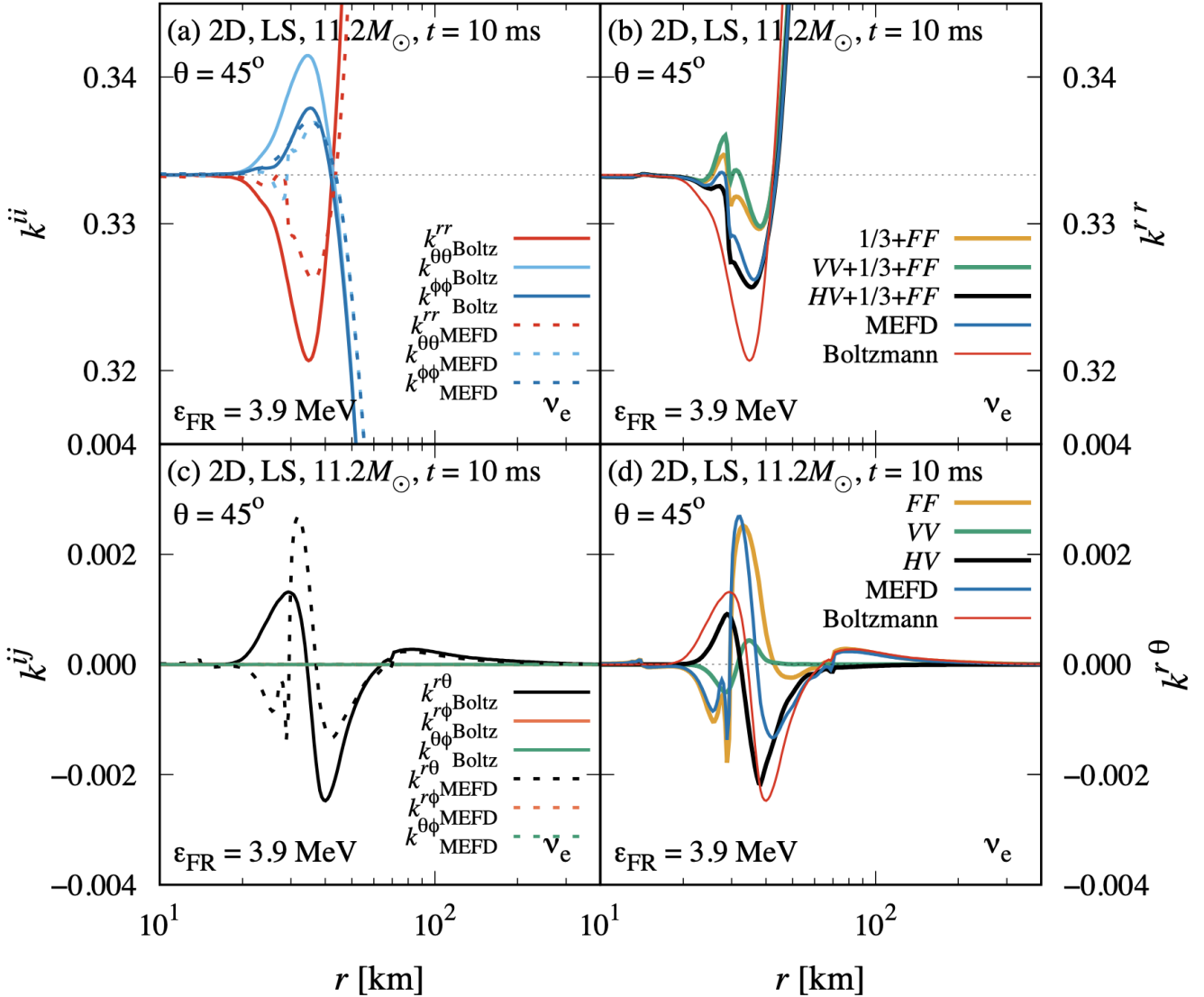}
\end{center}
\caption{Radial profiles of the diagonal and off-diagonal components (left panels) and the decomposed components (right panels) of the Eddington tensor of $\nu_e$ for the non-rotating model at $\epsilon_\mathrm{FR}=3.9$ MeV, where ``1/3", ``VV", and ``HV" denote the 1st, 2nd, and 3rd terms of Eq.~(\ref{eq:Pthick}) multiplied by $3(1-p_\mathrm{MEFD})/2$, respectively, and ``FF" is the term of Eq.~(\ref{eq:Pthin}) multiplied by $(3p_\mathrm{MEFD}-1)/2$.
The reference frame is the LB.
\label{fig:Ed}}
\end{figure*}

\begin{figure*}[ht!]
\begin{center}
\includegraphics[width=\hsize]{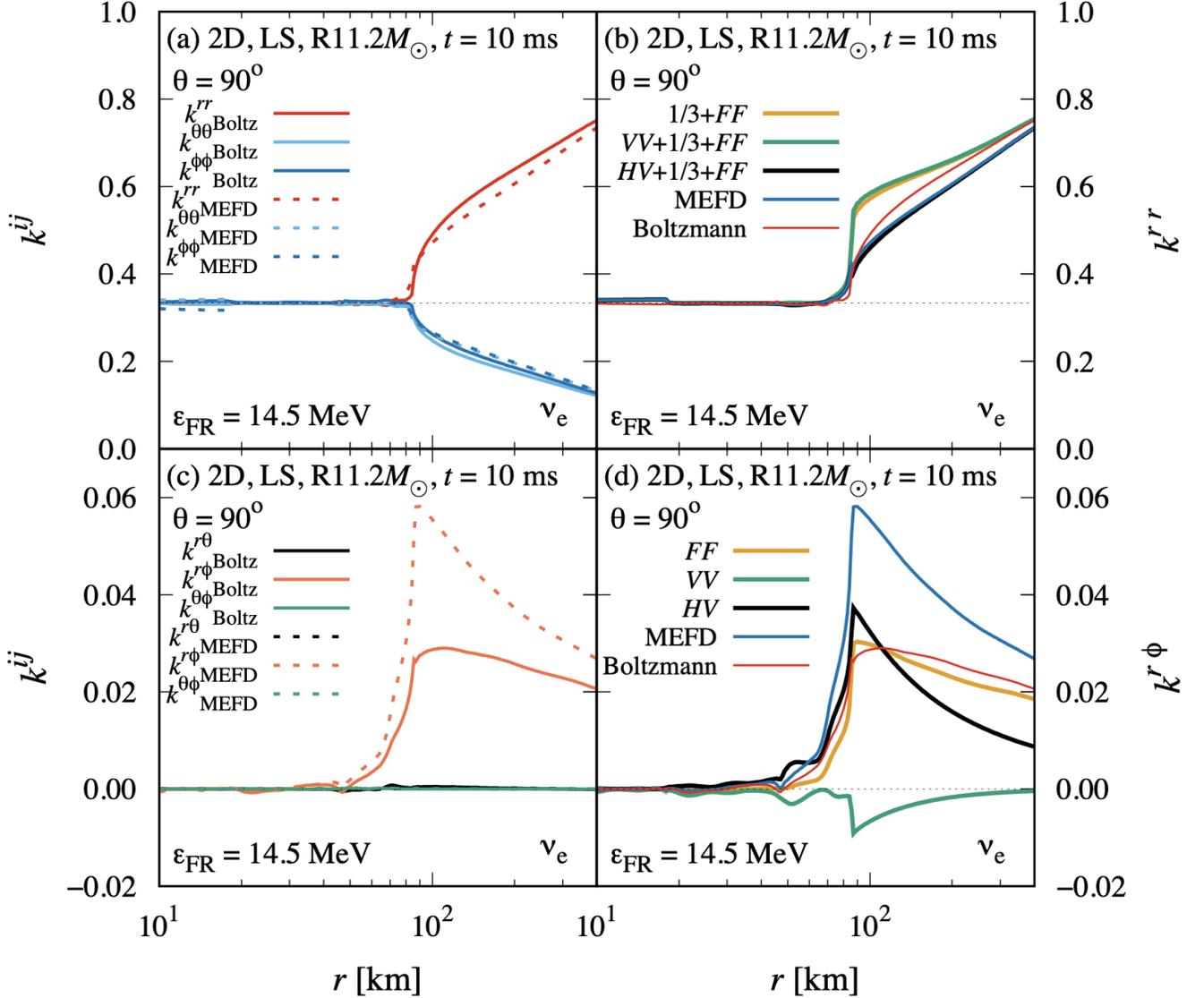}
\end{center}
\caption{As in Figure \ref{fig:Ed} but for the rotating model at $\epsilon_\mathrm{FR}=14.5$ MeV.
\label{fig:Ed2}}
\end{figure*}

Now we move to the main subject of this paper, i.e., the assessment of the closure relation under rapid rotation in 2D.
Although most of previous works \citep{Richers2017, Nagakura2018, Harada2019, Harada2020, Iwakami2020} studied the Levermore closure, we pick up the MEFD closure, since we have found in the preceding section that it performs better in 1D than others including the Levermore closure at least in the current context.
Instead of directly comparing the Eddington tensors as in other works \citep{Richers2017, Nagakura2018, Harada2020}, we compare its principal axes in this paper.
Since the Eddington tensor is a second-rank symmetric tensor, all of its three eigenvalues are real and the corresponding eigenvectors are orthogonal to one another, characterizing its anisotropy.
In fact, one of the remarkable properties that characterize the truncated moment method is that they employ the flux vector alone (in addition to Kronecker's delta) to build the Eddington tensor in the FR; as a result, the eigenvectors are the flux vector itself and two other mutually-orthogonal vectors that are perpendicular to the flux.
The latter fact implies in turn that the longest principal axis is always parallel or transverse to the flux vector.
This is true in 1D but is not the case in multi-D in general.
In this paper, we investigate quantitatively to what extent and in which direction the longest principal axis is misaligned with the flux.
This has never been done before for rapidly rotating models; in fact, \citet{Harada2019} employed a slowly rotating model and paid attention only to the length of the longest principal axis as a function of the radius as in Fig.~\ref{fig:pfr1d} whereas \citet{Iwakami2020} studied in their 3D simulation the effect of convection on the direction of the principal axes of the Eddington tensor in the LB but rotation is entirely ignored.
It should be also mentioned that the frame, in which the closure relation is imposed, matters, since the closure relations in the LB are sometimes not exactly equivalent to those in the FR (see Eqs.~(\ref{eq:Pijm1})-(\ref{eq:Pthin}) above).
We hence repeat the same exploration in the LB and compare the results with those obtained in the FR.
This is also the first-ever attempt of the kind.

\subsection{Basic Features in Dynamics and Neutrino Emissions}

This section summarizes briefly the basic features of the 2D-simulation results for the 11.2$M_\odot$ and 15.0$M_\odot$ progenitor models.

Figures \ref{fig:EVNF2D11_2M10ms} and \ref{fig:EVNF2D15_0M10ms} show color maps of entropy with the matter velocity vectors (upper panels) and the number density of $\nu_\mathrm{e}$ with its average velocity vectors (lower panels) at $t=10$ ms for the 11.2$M_\odot$ and 15.0$M_\odot$ models, respectively. 
The average neutrino velocity $V_{\nu i}$ is defined as
\begin{eqnarray}
V_{\nu i}=\frac{\mathcal{G}_{i}}{\mathcal{N}}, \label{eq:vi}
\end{eqnarray}
where the neutrino number density $\mathcal{N}$ and number flux $\mathcal{G}_{i}$ are expressed as
\begin{eqnarray}
\mathcal{N} = \iint \ \mathcal{F} (\epsilon, \Omega_\nu) \ \epsilon^2 d\epsilon d\Omega_\nu, \label{eq:N} \\
\mathcal{G}_{i} =  \iint \  \hat{\ell}_{i}\ \mathcal{F} (\epsilon, \Omega_\nu) \   \epsilon^2 d\epsilon d\Omega_\nu, \label{eq:Fi}
\end{eqnarray}
with $\hat{\ell}_{i}$ being the unit vector in momentum space.
Integration over the neutrino energy and solid angle is done in the LB here.
The non-rotating models are presented in the left panels, and the rotating ones are shown in the right panels.
The shock wave corresponds to the boundary between orange and purple colors in the upper figures.
The prompt convection grows in the central region (Figs.~\ref{fig:EVNF2D11_2M10ms}(a), \ref{fig:EVNF2D11_2M10ms}(b), \ref{fig:EVNF2D15_0M10ms}(a), and \ref{fig:EVNF2D15_0M10ms}(b)), and neutrinos move in various directions with matter for both non-rotating and rotating models (Figs.~\ref{fig:EVNF2D11_2M10ms}(c), \ref{fig:EVNF2D11_2M10ms}(d), \ref{fig:EVNF2D15_0M10ms}(c), and \ref{fig:EVNF2D15_0M10ms}(d)).
The anisotropy of the shock-wave geometry and neutrino propagation in the outer region is weak for the non-rotating models, and strong for the rotating models.

Figure \ref{fig:hydrovel2d10ms} shows the radial profiles of the velocity components $V_{r}$, $V_{\theta}$, and $V_{\phi}$ for the $\theta=\pi/2$ and $\pi/4$ directions, where $V_{r}=v^r$, $V_{\theta}=r v^\theta$, and $V_{\phi}=r\sin\theta\ v^\phi$.
Figure \ref{fig:hydrorho2d10ms} presents the radial profiles of the density $\rho$ and the mean free path divided by the radius $\Lambda$ for the $\theta=0$ and $\pi/2$ directions.
The results for the 11.2$M_\odot$ and 15.0$M_\odot$ models are plotted in the left and right panels, respectively.
Although the 11.2$M_\odot$ model experiences stronger prompt convection than the 15.0$M_\odot$ model,
the basic features are very similar between the models at $t=10$ ms for $r \lesssim 100$ km (Figs.~\ref{fig:hydrovel2d10ms} and \ref{fig:hydrorho2d10ms}).
For the non-rotating models, the radial distributions of $\rho$ and $\Lambda$ in the equatorial and polar directions agree well with each other.
By contrast, for the rotating models, the shock wave in the equatorial direction propagates faster than in the polar direction, and the resulting radial profiles of $\rho$ and $\Lambda$ in each direction are different (Fig.~\ref{fig:hydrorho2d10ms}).
Since such characteristics are common between the models, the results of the analysis for the neutrino transport at $t=10$ ms are essentially the same between the models.
Thus, only the results for the 11.2$M_\odot$ model are shown in the following sections.

\subsection{Principal-Axis Analysis in the Fluid-Rest Frame}

This section presents the results of the principal-axis analysis in the FR for both the non-rotating and rotating progenitor models, the main results of this paper.
In their neutrino-radiation hydrodynamics codes, \citet{Just2015} and \citet{Skinner2019} solved the two-moment equations in the FR with the closure relation imposed in this frame, whereas \citet{OConnor2015} employed the moment equations in the LB but applied the closure relations in the FR by consistently transforming the neutrino stress energy tensor obtained in the LB.
\citet{Just2015} compared the results for the Levermore and MEFD closures and found essentially the identical results for the time evolutions of the shock wave, neutrino luminosities and mean energies.
Note, however, that neither of the two closures can reproduce $p < 1/3$ at $e < 0.5$.
We compare the results for the Boltzmann simulation with those for the MEFD closure, paying particular attention to the misalignment of the longest principle axis with the flux vector in the following.

Figure~\ref{fig:INFR2DLS11_2en004} shows the color maps of $|\mu_{FL}|$ in the FR for $\nu_e$ at $t=10$ ms for $\epsilon_\mathrm{FR}=3.9$ MeV, where $\mu_{FL} = \cos\theta_{FL}$ with $\theta_{FL}$ is the angle that the longest principal axis makes with the flux vector (see Fig.~\ref{fig:ellipsoid}).
The dark blue region is the location, where $L$ is perpendicular to $F$ and $p<1/3$ is commonly observed.
Since the central region at $r \lesssim 10$ km is highly isotropic, having anisotropies of $O(10^{-8})$, it is ignored in the following discussion.
For $\epsilon_\mathrm{FR}=3.9$ MeV, this blue regions is circular for the non-rotating model both in the Boltzmann and MEFD calculations (Figs.~\ref{fig:INFR2DLS11_2en004}(a) and (c)), and it is oblate for the rotating model in both cases (Figs.~\ref{fig:INFR2DLS11_2en004}(b) and (d)).
The radial position of $\Lambda \approx 1$ is shifted inward near the poles in the rotating model (see the light blue line in Fig.~\ref{fig:hydrorho2d10ms}(c)).
Although the extent of the blue regions are not much different between the Boltzmann and MEFD cases, the color depth tends to be lighter for the Boltzmann case (Figs.~\ref{fig:INFR2DLS11_2en004}(a) and (b)) than for the MEFD case (Figs.~\ref{fig:INFR2DLS11_2en004}(c) and (d)).
This is because the longest principal axis is perfectly perpendicular to the flux vector in the MEFD case while it is inclined by angles between 0 and $\pi/2$ in the Boltzmann case.
The convective and/or rotational matter motion makes the neutrino distribution non-axisymmetric around the flux direction in momentum space and, as a result, the longest principal axis of the Eddington tensor becomes tilted from the perpendicular direction to $F$.

Figure~\ref{fig:INFR2DLS11_2en009} is the same as the previous figure except for the neutrino energy $\epsilon_\mathrm{FR}=14.5$ MeV.
For the non-rotating model (left panels), two dark blue regions are observed for Boltzmann (Fig.~\ref{fig:INFR2DLS11_2en009}(a)), whereas there is only one domain exists for MEFD.
The inner region of the Boltzmann case partly overlaps with the uniform dark blue region for MEFD while the outer one just behind the shock wave does not appear for the MEFD closure (Fig.~\ref{fig:INFR2DLS11_2en009}(c)). 
More interestingly, in the rotating model, the inner dark blue region is not clearly observed in the Boltzmann case (Fig.~\ref{fig:INFR2DLS11_2en009}(b)) whereas it is still there, being slightly oblate, in the MEFD case.
This is likely because the neutrino distribution, which is otherwise almost isotropic in this region (see the red line in Fig.~\ref{fig:fnueang}(b)), becomes tilted in the azimuthal direction owing to the rotation.
The outer dark blue region also nearly disappears particularly in the vicinity of the poles (Fig.~\ref{fig:INFR2DLS11_2en009}(b)). 
This happens because the hemispheric distribution is disrupted by fast non-radial flows in these regions in the presence of rotation (Fig.~\ref{fig:EVNF2D11_2M10ms}(b)).
In the equatorial region, the blue color becomes lighter in the rotating model (Figs.~\ref{fig:INFR2DLS11_2en009}(b)).
This time the hemispheric distribution is deformed to non-axisymmetric ones because of the azimuthal matter motion and neutrino flux again induced by the rotation.

The distribution function $\mathcal{F}$ of $\nue$ in momentum space and the corresponding ellipsoid of the Eddington tensor, both in the FR, are presented in Figure~\ref{fig:NeutrinoEllipsoidFR}.
The hemispheric distribution, corresponding to red isosurfaces, is observed in momentum space at $r=30$ km for the non-rotating model, where the prompt convection grows at $t=10$ ms (Fig.~\ref{fig:NeutrinoEllipsoidFR}(a)).
Note that the hemisphere is slanted.
At $r=65$ km for the rotating model, the hemispheric distribution, denoted by bluish isosurfaces, also emerges in momentum space (Fig.~\ref{fig:NeutrinoEllipsoidFR}(b)).
Since the distribution is not completely axisymmetric with respect to $F$ owing to the rotation, the axis $L$ is not perfectly perpendicular to $F$ for Boltzmann (Figs.~\ref{fig:NeutrinoEllipsoidFR}(d) and (e)).
By contrast, $L$ is always exactly perpendicular or parallel to $F$ for MEFD (Figs.~\ref{fig:NeutrinoEllipsoidFR}(g) and (h)) by its assumption of axisymmetry in momentum space.
In the preshock region, say at $r=90$ km, where matter falls almost freely with supersonic speeds even in the rotating model, however, $L$ is almost completely parallel to $F$ for both Boltzmann and MEFD.
The forward-peaked distribution is simply formed in momentum space (Figs.~\ref{fig:NeutrinoEllipsoidFR}(c), (f), and (i)).
These results indicate that rapid non-radial matter motions induced either by convection or by rotaion can easily disrupt axisymmetry  of the neutrino distribution in momentum space and make the longest principal axis neither perpendicular nor parallel to the flux vector.
The rapid rotation, in particular, slants the ellipse of the Eddington tensor in the $\phi$ direction.

\subsection{Principal-Axis Analysis in the Laboratory Frame}

So far we have been working in the fluid rest frame.
The assessment of the closure relations in the laboratory frame may be also important, though, since the closure relations are sometimes applied in this frame.
In fact, \citet{Shibata2011} gave the closure relation written with the radiation pressure tensor in the LB, and some previous investigations \citep{Richers2017, Nagakura2018, Harada2019, Harada2020, Iwakami2020} were conducted in the LB.
Picking up a snapshot at 100ms after bounce obtained by a Boltzmann simulation and comparing $P^{rr}, P^{\theta\theta}$, and $P^{r\theta}$ derived in the simulation and those calculated with the Levermore closure, \citet{Richers2017} showed that the relative error in the off-diagonal component is larger than that in the diagonal component in the region, where proto-neutron star convection is developed.
\citet{Nagakura2018} also pointed out that the difference in $k^{r\theta}$ between the Boltzmann transport and the Levermore closure they observed in the semitransparent region at $t=15$ ms after bounce is intrinsic and never reduced.
\citet{Harada2020}, on the other hand, argued that larger $k^{r\theta}$ for the Levermore closure is due to significant lateral motions caused by the prompt convection.

The effect of rotation was considered by \citet{Harada2019}, who investigated the Eddington tensor at $t=12$ ms after bounce for their slowly rotating model.
They first paid attention to the individual off-diagonal components of the Eddington tensor as in the previous studies.
Since their model was a slow rotator, they gave a detailed analysis only to $k^{r\theta}$ although rotation should affect $k^{r\phi}$ more directly.
They proceeded then to the analysis of the principal-axes of the Eddington tensor.
After just mentioning that they do not coincide with the $r$-, $\theta$-, and $\phi$-axes, since all off-diagonal components are nonvanishing, they focused on the radial profiles of the three eigenvalues of the Eddington tensor in their assessment of the closure relations.
As we mentioned earlier, the analysis of the eigenvalues alone are insufficient in multi-dimensions and the corresponding eigenvectors should be also studied, which we have done in the fluid rest frame in the preceding section.
As the closure relations are not always exactly equivalent between the two frames (see Eqs.~(\ref{eq:Pijm1_FR})-(\ref{eq:Pthin_FR}) and Eqs.~(\ref{eq:Pijm1})-(\ref{eq:Pthin})), it is non-trivial if the finding in the FR is also true in the LB.
Note in passing that the principal-axis analysis, considering both eigenvalues and eigenvectors, was applied already by \citet{Iwakami2020} to their 3D model.
It is non-rotating, though.
The investigation here is hence complementary to all the previous studies one way or another.
We also address the systematic difference between the frames.
The individual components of the Eddington tensor are also inspected to understand the effect of convection and rotation on $k^{r\theta}$ and $k^{r\phi}$, respectively.

Figures~\ref{fig:INLB2DLS11_2en004} and \ref{fig:INLB2DLS11_2en009} show the color maps of $|\mu_\mathrm{FL}|$ in the LB for $\nu_e$ at $t=10$ ms for $\epsilon_\mathrm{FR}=3.9$ and 14.5 MeV, respectively.
For Boltzmann transport, the blue colored region in the LB (Figs.~\ref{fig:INLB2DLS11_2en004}(a), \ref{fig:INLB2DLS11_2en004}(b), \ref{fig:INLB2DLS11_2en009}(a), and \ref{fig:INLB2DLS11_2en009}(b)) appears at a similar position to the one in the FR (Figs.~\ref{fig:INFR2DLS11_2en004}(a), \ref{fig:INFR2DLS11_2en004}(b), \ref{fig:INFR2DLS11_2en009}(a), and \ref{fig:INFR2DLS11_2en009}(b)).
However, the color depth for the LB is lighter than that for the FR in the regions, where fast non-radial motions exist.
Such light blue regions are sporadical in the non-rotating model (Fig.~\ref{fig:INLB2DLS11_2en004}(a)) whereas they are much more extended in the rotating model (Figs.~\ref{fig:INLB2DLS11_2en004}(b) and \ref{fig:INLB2DLS11_2en009}(b)).
The corresponding distribution functions $\mathcal{F}$ of $\nue$ in momentum space and the ellipsoids of the Eddington tensor in the LB are presented in Figs.~\ref{fig:NeutrinoEllipsoidLB}(a) and (d) for the non-rotating model and in Figs.~\ref{fig:NeutrinoEllipsoidLB}(b) and (e) for the rotating model.
When the matter radiating neutrinos moves, the direction of neutrino propagation approaches the direction of the motion owing to the relativistic beaming effect \citep{Nagakura2014}.
In the presence of rotation, the ellipse is inclined not only in the $\theta$ direction but also in the $\phi$ direction, which implies in turn that neutrinos carry angular momentum \citep{Harada2019}.
The effect of the non-radial motion of matter on the hemispheric distribution of neutrinos in momentum space is opposite to the effect on the forward-peaked distribution: it diverts $L$ from $F$ for the former (Figs.~\ref{fig:NeutrinoEllipsoidLB}(d), (e)) while it makes $L$ closer to $F$ for the latter (Fig.~\ref{fig:NeutrinoEllipsoidLB}(f)).

The results of the MEFD closure for the LB change from those for the FR but in a different manner from what we have observed above for the Boltzmann transport.
In fact, for the non-rotating model, light blue regions appear around the central dark blue region for $\epsilon_\mathrm{FR}=3.9$ MeV (Figs.~\ref{fig:INLB2DLS11_2en004}(c) and (d)); the central dark blue region is shrunk for $\epsilon_\mathrm{FR}=$14.5 MeV (Figs.~\ref{fig:INLB2DLS11_2en009}(c) and (d)).
For the rotating model, on the other hand, greenish regions emerge around the central dark blue region (Figs.~\ref{fig:INLB2DLS11_2en004}(d) and  \ref{fig:INLB2DLS11_2en009}(d)).
In these regions, $F$ and $L$ are misaligned with each other in the LB (Figs.~\ref{fig:NeutrinoEllipsoidLB}(f) and (i)) while they are well aligned in the FR (Figs.~\ref{fig:NeutrinoEllipsoidFR}(f) and (i)).
This is understandable from Eqs.~(\ref{eq:Pthick}): in the LB, the Eddington tensor is constructed not only from the flux vector but also from the matter velocity.
It is interesting to point out that the angle between $F$ and $L$ for the MEFD closure is larger than that for the Boltzmann transport at $r=90$ km, i.e., in the preshock region, in the LB.

The difference in $\theta_{FL}$ between the Boltzmann calculation and the MEFD-closure approximation for the rotating model comes from the $k^{r\phi}$ component in the LB.
To see this, we plot in the left panels of Figures~\ref{fig:Ed} and \ref{fig:Ed2} the radial distributions of diagonal (upper panels) and off-diagonal (lower panels) components of the Eddington tensor for $\nu_e$ at $t=10$ ms for the non-rotating and rotating models, respectively.
The Eddington tensors $k^{ij}_\mathrm{Boltz}$ and $k^{ij}_\mathrm{MEFD}$ are calculated directly and in the MEFD-closure approximation, respectively, from the numerical data obtained by the Boltzmann simulation.
In the non-rotating model,  the term $k^{\theta\theta}_\mathrm{MEFD}$ is smaller than  $k^{\phi\phi}_\mathrm{MEFD}$ for $\epsilon_\mathrm{FR}=3.9$ MeV at $r=30$ km, while $k^{\theta\theta}_\mathrm{Boltz}$ is larger than $k^{\phi\phi}_\mathrm{Boltz}$ at the same position (Fig.~\ref{fig:Ed}(a)).
This is why the longest axis of the ellipsoid for MEFD is perpendicular to the $P_r-P_\theta$ plane at this radius in the LB (Fig.~\ref{fig:NeutrinoEllipsoidLB}(g)).
The off-diagonal component $k^{r\theta}$ is non-vanishing (Fig.~\ref{fig:Ed}(c)) owing to the prompt convection for the non-rotating model and tilts the longest principal axis of the ellipsoid from the coordinate axes (Fig.~\ref{fig:NeutrinoEllipsoidLB}(d)).
The radial profile is qualitatively different between the Boltzmann calculation and the MEFD-closure approximatioin.
For the rotating model, on the other hand, $k^{r\phi}$ grows steeply with radius at $r\lesssim 90$ km and then decreases gradually at $r\gtrsim 90$ km for $\epsilon_\mathrm{FR} =14.5$ MeV (Fig.~\ref{fig:Ed2}(c)).
The misalignment of $L$ and $F$ is related to this growth of $k^{r\phi}$ (Figs.~\ref{fig:NeutrinoEllipsoidLB}(f) and (i)).
It should be pointed out that the MEFD closure overestimates the value of $k^{r\phi}$ by a factor of 2 (Fig.~\ref{fig:Ed2}(c)).
Although it is not shown in this paper, this is also found for the Levermore closure, which is not observed in the previous papers for the slowly rotating model \citep{Harada2019} and for the 3D non-rotating model \citep{Iwakami2020}.

To identify the origin of the discrepancy between $k^{r\phi}_\mathrm{MEFD}$ and $k^{r\phi}_\mathrm{Boltz}$ in the LB, the contributions of individual terms in Eq.~(\ref{eq:Pijm1})$-$(\ref{eq:Pthin}) are investigated.
The right panels of Figures \ref{fig:Ed} and \ref{fig:Ed2} present the decompositions of the diagonal (upper panels) and off-diagonal (lower panels) components of the Eddington tensor in the LB for the non-rotating model at $\epsilon_\mathrm{FR}=3.9$ MeV and for the rotating model at $\epsilon_\mathrm{FR}=14.5$ MeV, where ``1/3", ``VV", and ``HV", respectively, denote the 1st, 2nd, and 3rd terms of Eq.~(\ref{eq:Pthick}) multiplied by $3(1-p_\mathrm{MEFD})/2$, and ``FF" is the right hand side of Eq.~(\ref{eq:Pthin}) multiplied by $(3p_\mathrm{MEFD}-1)/2$.
In the diagonal components of the Eddington tensor, the terms ``1/3" and ``FF" are the terms common to the FR and the LB.
The term ``HV", on the other hand, plays an important role pushing $k^{rr}_\mathrm{MEFD}$ close to $k^{rr}_\mathrm{Boltz}$ in the region of $\Lambda \lesssim 10$ (Figs.~\ref{fig:hydrorho2d10ms}(c) and \ref{fig:Ed}(b)).
In the off-diagonal component, the term ``HV" alone reproduces the results for Boltzmann rather well, and the term ``FF" seems to be responsible for the discrepancy we observed at $\Lambda \lesssim 10$ (Fig.~\ref{fig:Ed}(d)).
At $\Lambda \gtrsim 10$ (Fig.~\ref{fig:hydrorho2d10ms}(e)), however, the roles of the terms ``FF" and ``HV" are inverted (Figs.~\ref{fig:Ed2}(b) and \ref{fig:Ed2}(d)).
This indicates that the interpolation of the optically thick and thin limits employed in Eq.~(\ref{eq:Pijm1}) does not perform very well in the transition region in the LB.
Note that in the derivation of the pressure tensor in the LB, in which the terms ``HV" and ``VV" obtained, the equations are truncated at the first order of the mean free path \citep{Shibata2011}. 
Hence, higher-order terms may be needed to better reconstruct the pressure tensor in the transition region in the LB \citep[see also][]{Harada2019, Harada2020}.
Although not presented here, we confirmed that this is a common problem to the Levermore closures in Eq.~(\ref{eq:p_Levermore}).

\section{CONCLUSIONS \label{sec:conclusion}}

Using the data at the early post bounce phase derived by the 2D full Boltzmann neutrino transport simulations of core-collapse supernovae for 11.2$M_\odot$ and 15.0$M_\odot$ progenitor models with and without rotation, we have assessed the performance of some closure relations that are used commonly in the truncated moment method for neutrino transport in the literature.
We have first compared the Eddington factors given by these closure relations with the results of the 1D simulations in spherical symmetry for the non-rotating models.
We have then applied the principal axis analysis to the results of the 2D simulations in axisymmetry for both the non-rotating and rotating models.
We have studied in detail the eigenvectors of the Eddington tensor and made a comparison between the simulation results and the maximum entropy closure for Fermi-Dirac radiations (MEFD).
The new findings of this study are summarized as follows.

\begin{enumerate}
\item
The comparison of the Eddington factor obtained directly from the 1D simulation results and the Eddington factors estimated via the closure relations from the energy density and flux in the same simulation results confirmed that MEFD performs best among the closures investigated but that even the MEFD closure fails when $p<1/3$ happens simultaneously with $e < 0.5$, which occurs, for example, immediately behind the shock wave in the early post bounce phase.

\item
The application of the principal axis analysis to the Eddington tensors that are obtained either directly from the 2D simulation results or via the MEFD closure in the fluid rest frame (FR) has revealed that the longest principal axis $L$ in the simulation results is neither parallel nor perpendicular to the flux vector $F$.
This is in sharp contrast to the estimates by the MEFD closure, in which $L$ is always either parallel or perpendicular to $F$ because of the axisymmetry assumed in momentum space.
The difference is particularly remarkable in the region of rapid convection and/or rotation.
In particular, the rapid rotation makes the ellipse of the Eddington tensor inclined in the $\phi$ direction.

\item
The same analysis has been repeated in the laboratory frame (LB).
Since the closure relations are not always exactly equivalent between the two frames as in Eqs.~(\ref{eq:Pijm1_FR})-(\ref{eq:Pthin_FR}) and Eqs.~(\ref{eq:Pijm1})-(\ref{eq:Pthin}), this is non-trivial.
We have observed that $L$ is neither parallel nor perpendicular to $F$ even for the MEFD closure as it includes explicitly velocity-dependent terms in the LB.
We have demonstrated, however, that the angle that $L$ makes with $F$ is different between the direct calculations and the estimates by the closure relation.
In fact, the radial profiles of the $k^{r\theta}$ are qualitatively different between the two cases while $k^{r\phi}$ is overestimated by the MEFD closure by a factor of two for the rotating model.
We have also found that these discrepancies, which are most remarkable in the regions where matter moves non-radially by convection and/or rotation, are caused by inaccuracy of the interpolation between the optically thick and thin limits.
These results are valid not only for the MEFD closure but also for the Levermore closure.
\end{enumerate}

The last point may be the most important, since the closure relation is normally imposed in the laboratory frame in the truncated moment methods in the literature.
Now that there have been many 3D simulations of CCSNe conducted with the truncated moment methods with some closure relation and their results seem to be converging among different groups, we believe that the subtleties in the closure relations, particularly when it is employed in the laboratory frame, should be taken seriously and studied quantitatively. 

Of course we need to extend the analysis to later phases, where the neutrino-driven convection and/or the standing accretion shock instability (SASI) are in operation in the gain region, and the lepton-driven convection is developing in the proto-neutron star. 
In so doing, higher resolutions should be used for both space and momentum space to reproduce forward-peaked angular distributions accurately.
Such simulations will be done on the Fugaku supercomputer in Japan.
Not to mention, 3D rotating models will be also investigated.

\begin{acknowledgments}
This research used the K and Fugaku supercomputers and the high-performance computing resources of FX100 at Nagoya University ICTS, Grand Chariot at Hokkaido University, and Oakforest-PACS at JCAHPC through the HPCI System Research Project (Project ID: hp130025, 140211, 150225, 150262, 160071, 160211, 170031, 170230, 170304, 180111, 180179, 180239, 190100, 190160, 200102, 200124, 210050, 210051, 210164).
The new system Supercomputer ``Flow" at Nagoya University ICTS, NEC SX Aurora Tsubasa at KEK, the Research Center for Nuclear Physics (RCNP) at Osaka University, and XC50 and the general common-use computer system operated by CfCA at the National Astronomical Observatory of Japan (NAOJ) also contribute to this study.
Large-scale storage of numerical data is provided by Japan Lattice Data Grid (JLDG). 
This work was supported in part by Grants-in-Aid for Scientific Research (26104006, 15K05093, and 19K03837) and the Grant-in-Aid for Scientific Research on Innovative areas ``Gravitational wave physics and astronomy: Genesis” (17H06357 and 17H06365) and ”Unraveling the History of the Universe and Matter Evolution with Underground Physics” (19H05802 and 19H05811) from the Ministry of Education, Culture, Sports, Science and Technology (MEXT), Japan. This work was also partly supported by research programs at the K-computer of the RIKEN R-CCS, HPCI Strategic Program of Japanese MEXT, Priority Issue on Post-K-computer (Elucidation of the Fundamental Laws and Evolution of the Universe), the Joint Institute for Computational Fundamental Sciences (JICFus), and the Japan Society for the Promotion of Science (JSPS) Grant-in-Aid for Young Scientists(Start-up, JP19K23435).
A. H. was supported in part by MEXT Grant-in-Aid for Research Activity Start-up (19K23435).
S. F. was supported by JSPS KAKENHI (19K14723).
S. Y. is supported by the Institute for Advanced Theoretical and Experimental Physics, and Waseda University and the Waseda University Grant for Special Research Projects (project number: 2020-C273).
\end{acknowledgments}

\appendix

\section{Resolution test \label{sec:resolution}}

\begin{figure}[ht!]
\begin{center}
\includegraphics[width=\hsize]{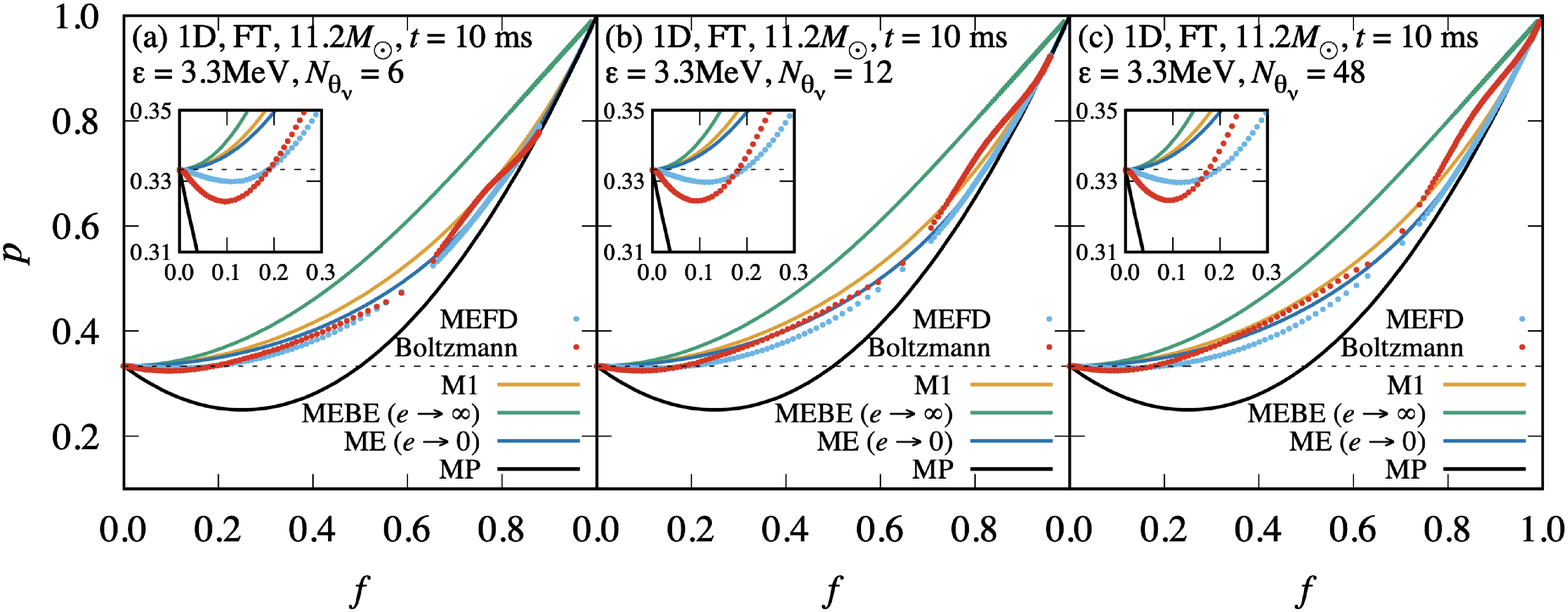}
\end{center}
\caption{The Eddington factor $p$ as a function of the flux factor $f$ for $\epsilon_{FR}$=3.9 MeV at $t=10$ ms for (a) $N_{\theta_\nu}=6$, (b) $N_{\theta_\nu}=12$, and (c) $N_{\theta_\nu}=48$. The Eddington factors were obtained by Boltzmann transport (red dots) and MEFD closure (light blue dots). The solid lines denote the closures of Levermore (yellow), MEBE in the limit of $e\rightarrow \infty$ (green), ME in the classical limit of $e\rightarrow 0$ (blue), and MP (black).
Note that the Furusawa--Togashi (FT) EOS is used in this test.
The neutrino energy in the range of 0 to 300MeV is divided into 16 bins. 
\label{fig:pfr1Den003}}

\begin{center}
\includegraphics[width=\hsize]{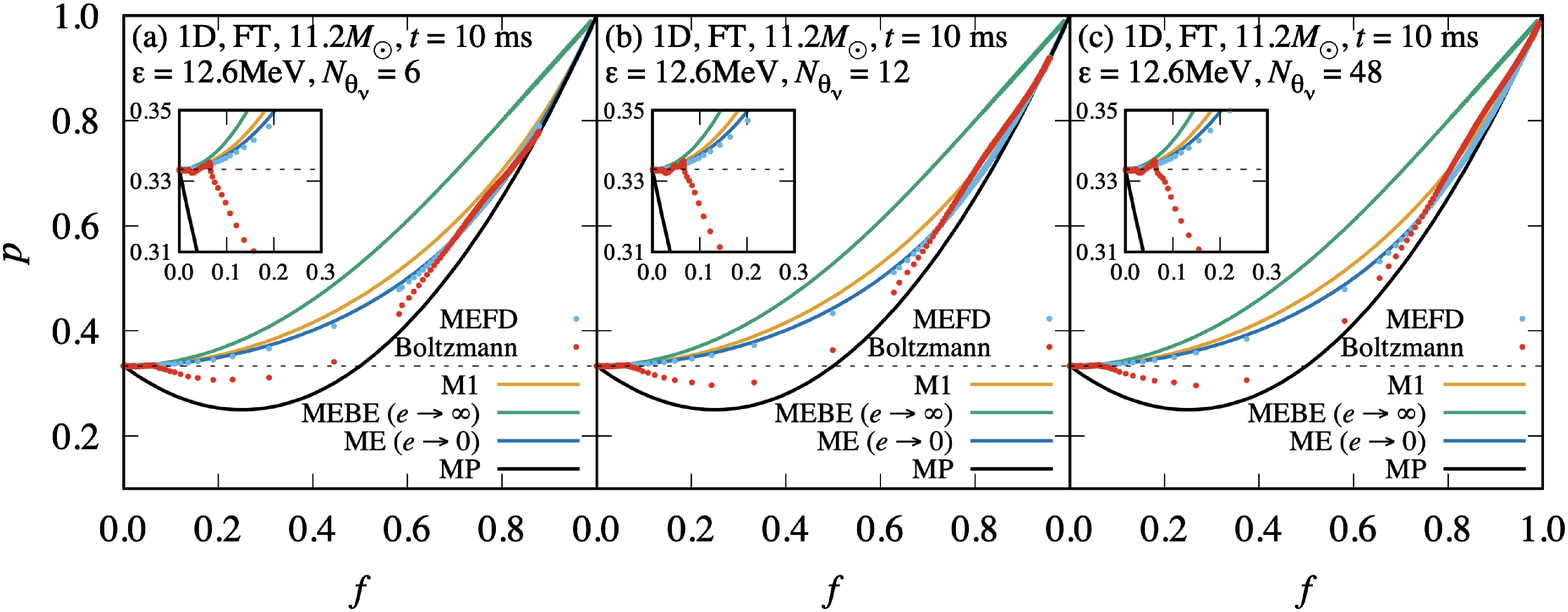}
\end{center}
\caption{As in Figure \ref{fig:pfr1Den003} but for $\epsilon_{FR}$=12.6 MeV, which roughly agrees with the average energy just behind the shock wave.
\label{fig:pfr1Den007}}
\end{figure}

This appendix shows the results of the resolution tests in this paper. 
Readers may also refer to earlier publications for more tests \citep{Richers2017, Nagakura2018, Harada2019, Iwakami2020}.
Using the data of the 1D resolution test in \citet{Iwakami2020}, we show the effect of the angular resolution on the Eddington factor $p$ as a function of the flux factor $f$.
In this test, the hydrodynamic simulations with Boltzmann transport start from the onset of core collapse after changing angular resolution.
There are some differences in the simulations between the main results and resolution tests in this paper.
The Furusawa--Togashi EOS \citep{Furusawa2017} is used in this test.
The neutrino energy mesh is divided into 16 bins in the resolution tests and into 20 bins in the main results.
Although the grid number of the polar angle in momentum space is $N_{\theta_\nu}=10$ in the main results, $N_{\theta_\nu}=6, 12$ and 48 are selected in the resolution tests.

Figures~\ref{fig:pfr1Den003} and \ref{fig:pfr1Den007} show $p-f$ plots for $\epsilon_\mathrm{FR}=3.3$ and 12.6 MeV, respectively.
The red dots denote $p_\mathrm{Boltzmann}$ and the light blue dots denote $p_\mathrm{MEFD}$.
The regions of $p < 1/3$ are roughly the same among $N_{\theta_\nu}=6$, 12, and 48.
The advantage of MEFD closure in the region of $p<1/3$ where $e>0.5$ is independent of the angular resolution.
However, the difference in $p_\mathrm{Boltzmann}$ among $N_{\theta_\nu}=6$, 12, and 48 increases as $f$ increases, especially for $\epsilon_\mathrm{FR} = 3.3$ MeV.
For Boltzmann transport simulations, the neutrinos tend to have a more forward-peaked distribution with increasing $N_{\theta_\nu}$.
Although the convergence is not completely achieved even in $N_{\theta_\nu}=48$, it was found that the closure matching with the results for Boltzmann transport depends on the flux factor $f$ and the neutrino energy $\epsilon$.
Hence, it is difficult to determine the best ``algebraic Eddington factor method" proposed so far.


\bibliography{library}{}
\bibliographystyle{aasjournal}



\end{document}